\documentclass{appolb}
\usepackage{epsfig}
\usepackage{rotating}
\usepackage{amsmath}
\usepackage{psfrag}
\allowdisplaybreaks[1]

\usepackage{subfigure}
\renewcommand{\thesubfigure}{(\arabic{subfigure})}

\def\theequation{\thesection.\arabic{equation}}
\newcommand{\RE}{{\rm Re}}
\newcommand{\IM}{{\rm Im}}
\newcommand{\vcb}{|V_{cb}|}
\newcommand{\vtd}{|V_{td}|}
\newcommand{\vub}{|V_{ub}/V_{cb}|}
\newcommand{\vts}{|V_{ts}|}
\newcommand{\vus}{|V_{us}|}

\def\R1{\varepsilon_1}
\def\E8{\varepsilon_8}

\def\epe{\varepsilon'/\varepsilon}
\def\as{\alpha_s}

\newcommand{\mt}{m_{\rm t}}
\newcommand{\mtb}{\overline{m}_{\rm t}}

\newcommand{\mb}{m_{\rm b}}
\newcommand{\mw}{M_{\rm W}}
\newcommand{\mz}{M_{\rm Z}}
\newcommand{\gev}{\, {\rm GeV}}

\newcommand{\mev}{\, {\rm MeV}}

\newcommand{\bea}{\begin{eqnarray}}
\newcommand{\eea}{\end{eqnarray}}
\newcommand{\bd}{\begin{displaymath}}
\newcommand{\ed}{\end{displaymath}}

\newcommand{\beq}{\begin{equation}}
\newcommand{\eeq}{\end{equation}}
\newcommand{\be}{\begin{equation}}
\newcommand{\ee}{\end{equation}}
\newcommand{\bi}{\begin{itemize}}
\newcommand{\ei}{\end{itemize}}
\newcommand{\ord}{{\cal O}}

\newcommand{\Ctilde}{\tilde{C}}
\def\kpn{K^+\rightarrow\pi^+\nu\bar\nu}
\def\kpnn{K^+\rightarrow\pi^+\nu\bar\nu}
\def\klpn{K_{\rm L}\rightarrow\pi^0\nu\bar\nu}
\newcommand{\kmm}{K_{\rm L} \to \mu^+ \mu^-}
\newcommand{\kpe}{K_{\rm L} \to \pi^0 e^+ e^-}

\newcommand{\relt}{\RE\lambda_t}
\newcommand{\relc}{\RE\lambda_c}

\headtitle{MINIMAL FLAVOUR VIOLATION}
\headauthor{Andrzej J. Buras}
\begin{document}

%%%%%%%%%%%%%%%%%%%%%%%%%%%%%%%%%

\thispagestyle{empty}
\phantom{xxx}
\vskip1truecm
\begin{flushright}
 TUM-HEP-530/03 \\
October 2003
\end{flushright}
\vskip1.0truecm
\centerline{\LARGE\bf MINIMAL FLAVOUR  VIOLATION}
   \vskip1truecm
\centerline{\Large\bf Andrzej J. Buras}
\bigskip
\centerline{\sl Technische Universit{\"a}t M{\"u}nchen}
\centerline{\sl Physik Department} 
\centerline{\sl D-85748 Garching, Germany}
\vskip1truecm
\centerline{\bf Abstract}

These lectures give a description of  models with 
minimal flavour violation (MFV) that can be tested in
$B$ and $K$ meson decays. 
This class of models can be formulated to a very good approximation 
in terms of 11 parameters: 4 parameters of the CKM matrix and 7 values 
of the {\it universal} master functions $F_r$ that parametrize 
the short distance 
contributions. In a given MFV model, $F_r$ can be calculated in 
perturbation theory and are generally correlated with each other but 
in a model independent analysis they must be considered as free 
parameters. We conjecture that only 5 or even 
only 4  of these 
functions receive significant new physics contributions. 
We summarize the status of the CKM matrix, 
outline strategies for the determination of the values of $F_r$ 
and present a number of relations between physical observables that do
not depend on $F_r$ at all.
We emphasize that the formulation of MFV in terms of master functions allows 
to study transparently correlations between $B$ and $K$ decays which 
is very difficult if 
Wilson coefficients normalized at low energy scales are used instead.
We discuss briefly a specific MFV model: the Standard 
Model with one 
universal large extra dimension. 

\vskip1truecm

\centerline{\it  Lectures given at }
\centerline{\bf the 43rd Cracow School for Theoretical Physics}
\centerline{\it Zakopane, May 30 -- June 06, 2003}
%%% end title page %%%%%%%%%%%%%

\newpage
\mbox{}
\thispagestyle{empty}
\newpage
\setcounter{page}{1}

%%%%%%%%%%%%%%%%%%%%%%%%%%%%%%%%%

%\phantom{xxx}
%\vskip1truecm
%\vskip1.0truecm
%\centerline{\LARGE\bf MINIMAL FLAVOUR VIOLATION}
%   \vskip1truecm
%\centerline{\Large\bf Andrzej J. Buras}
%\bigskip
%\centerline{\sl Technische Universit{\"a}t M{\"u}nchen}
%\centerline{\sl Physik Department} 
%\centerline{\sl D-85748 Garching, Germany}
%\vskip1truecm

\eqsec
\title{MINIMAL FLAVOUR  VIOLATION}
\author{Andrzej J. Buras
\address{Technische Universit{\"a}t M{\"u}nchen,
  Physik Department\\ D-85748 Garching, Germany}}
\maketitle

%\centerline{\bf Abstract}
\begin{abstract}
These lectures give a description of  models with 
minimal flavour violation (MFV) that can be tested in
$B$ and $K$ meson decays. 
This class of models can be formulated to a very good approximation 
in terms of 11 parameters: 4 parameters of the CKM matrix and 7 values 
of the {\it universal} master functions $F_r$ that parametrize 
the short distance 
contributions. In a given MFV model, $F_r$ can be calculated in 
perturbation theory and are generally correlated with each other but 
in a model independent analysis they must be considered as free 
parameters. We conjecture that only 5 or even 
only 4  of these 
functions receive significant new physics contributions. 
We summarize the status of the CKM matrix, 
outline strategies for the determination of the values of $F_r$ 
and present a number of relations between physical observables that do
not depend on $F_r$ at all.
We emphasize that the formulation of MFV in terms of master functions allows 
to study transparently correlations between $B$ and $K$ decays which 
is very difficult if 
Wilson coefficients normalized at low energy scales are used instead.
We discuss briefly a specific MFV model: the Standard 
Model with one 
universal large extra dimension. 
\end{abstract}

%%% MAIN TEXT

\section{Introduction}
The understanding of flavour dynamics is one of the most important goals of 
elementary particle physics. Because this understanding will likely come 
from very short distance scales, the loop induced processes like flavour 
changing neutral current (FCNC) transitions will for some time continue to 
play a crucial role in achieving this goal. They can be studied most
efficiently in 
$K$ and $B$ decays but  $D$ decays and hyperon decays can also offer 
useful information in this respect.

Within the Standard Model (SM), the FCNC processes are governed by
\begin{itemize}
\item
the unitary Cabibbo-Kobayashi-Maskawa (CKM) matrix \cite{CAB,KM}
that parametrizes the weak charged current interactions of quarks,
\item
the Glashow-Iliopoulos-Maiani (GIM) mechanism \cite{GIM} that forbids 
the appearance of FCNC 
processes at the tree level with the size of its violation at the one loop 
level depending sensitively on the CKM parameters and the masses of exchanged
particles,
\item
the asymptotic freedom of QCD \cite{ASF} that allows to calculate the 
impact of strong 
interactions on weak decays at sufficiently short distance scales within the 
framework of renormalization group improved perturbation theory,
\item
the operator product expansion (OPE) \cite{OPE} with
local operators having a specific Dirac structure and their matrix elements 
calculated by means of non--perturbative methods or in certain 
cases extracted from experimental data on leading decays with the help
of flavour symmetries.
\end{itemize}

The present data on rare and CP violating $K$ and $B$ decays are consistent 
with this structure but as only a handfull of FCNC processes have been 
measured, it is to be seen whether some modification of this picture will
be required in the future when the data improve.

In order to appreciate the simplicity of the structure of FCNC processes 
within the SM, let us realize that although the CKM matrix is introduced
in connection with charged current interactions of quarks, 
its departure from the unit matrix is the origin of all flavour violating 
and CP-violating transitions in this model. Out there, at very short 
distance scales, the picture could still be very different. In particular, 
new complex phases could be present in both charged and neutral tree 
level interactions, the GIM mechanism could be violated already at 
the tree level, 
the number of parameters describing flavour violations could be significantly
larger than the four present in the CKM matrix and the number of operators 
governing the decays could also be larger. We know all this through 
extensive studies of complicated extensions of the SM.

In these lectures we will discuss a class of models in which the general 
structure of FCNC processes present in the SM is preserved. In particular
all flavour violating and CP-violating 
transitions are governed by the CKM matrix and the only relevant local 
operators are the ones that are relevant in the SM.
We will call this scenario ``Minimal Flavour 
Violation" (MFV) \cite{UUT} being aware of the fact that for some authors 
\cite{AMGIISST,BOEWKRUR}
MFV means a more general framework in which also new operators can give 
significant contributions. 

In the MFV models, as defined in \cite{UUT},  the decay amplitudes for 
any decay of interest 
can be written as follows \cite{PBE,BH92}
\be\label{mmaster}
{A({\rm Decay})}= 
P_c({\rm Decay}) + \sum_r P_r({\rm Decay} ) \, F_r(v),
\end{equation}
with $F_r(v)$ being {\it real}. $P_c$  summarizes 
contributions stemming from light internal quarks,
in particular the charm quark, and the sum incorporates the remaining 
contributions.

The objects $P_c$, $P_r$ and $F_r$ have the following  general properties:
%\begin{itemize}
%\item

{\bf 1.} $F_r(v)$ are {\it process independent universal} 
``master functions" that in the 
SM reduce to 
the so--called Inami--Lim functions \cite{IL}. The master functions result
from the calculations of various
box and penguin diagrams. In the SM they depend only on the ratio 
$m_t^2/M_W^2$ but in other models new parameters enter,  
like the masses of charginos, 
squarks, charged Higgs particles and $\tan\beta$ in the MSSM and the 
compactification radius $R$ in models with large extra dimensions. 
We will collectively denote these parameters by $v$.
Up to some reservations to be made in the next section, there are {\it seven} 
master functions \cite{PBE,BH92} that in a given MFV model can be 
calculated as functions 
of $v$. As in some extensions of the SM the number of free parameters in 
$F_r$ is smaller than seven, the functions $F_r$ are not always independent 
of each other. However, in a general model independent analysis of MFV we 
have 
to deal with seven parameters, the values of the master functions, that can 
be in principle determined experimentally. Now, there are several decays 
to which a single master function contributes. These decays are 
particularly suited for the determination of the value of this single
function. Equally important, by taking ratios of branching ratios for 
different decays, it is possible to obtain relations between observables that
do not depend on $F_r(v)$ at all. These relations can be regarded as 
``sum rules" of MFV. Their violation by experimental data would indicate the 
presence of new complex phases beyond the CKM phase and/or new local 
operators and generally new sources of flavour and CP violation.
Finally, explicit calculations indicate that the number of relevant master
functions can be reduced to {\it five} or even {\it four} in which case 
the system becomes 
more constrained. We will discuss all this below.
%\item

{\bf 2.} The coefficients $P_c$ and $P_r$ are {\it process dependent}
but within the class of MFV models they are {\it model independent}. 
$P_c$ and $P_r$ depend on hadronic matrix elements of local operators $Q_i$ 
that usually are parametrized by $B_i$ factors. The latter can be calculated 
in QCD
by means of non--perturbative methods.
For instance in the case of 
$K^0-\bar K^0$ mixing, the matrix element of the operator
$\bar s \gamma_\mu(1-\gamma_5) d \otimes \bar s \gamma^\mu(1-\gamma_5) d $
is represented by the parameter $\hat B_K$.
There are other non-perturbative parameters in the MFV that represent 
matrix elements of operators $Q_i$ with different colour and Dirac 
structures. 
The important property of $P_c$ and $P_r$ is their manifest independence of
the choice of the operator basis in the effective weak Hamiltonian \cite{PBE}.
$P_c$ and $P_r$ include also QCD factors 
$\eta_i^{QCD}$ that summarize the renormalization group effects at scales
below $\mu=\ord(M_W,m_t)$. Similar to the $B_i$ factors, $\eta_i^{QCD}$ 
do not depend on a
particular MFV model and can be simply calculated within the SM. 
This universality of $\eta_i^{  QCD}$ requires a careful treatment of
QCD corrections at scales $\mu=\ord(M_W,m_t)$ as discussed in Section 2.
Finally,  
$P_c$ and $P_r$ depend on the four parameters of the CKM matrix. As the 
l.h.s of (\ref{mmaster}) can be extracted from experiment and is $v$ 
independent, while the r.h.s involves $v$ that are specific to a given 
MFV model, the values of the CKM parameters extracted from the data must 
in principle be $v$ dependent in order to cancel the $v$ dependence of 
the master 
functions. On the other hand, as stated in the first item, it is possible 
to consider ratios of branching ratios in which the functions $F_i$ cancel 
out. Such ratios are particularly suited for the determination of the true 
universal CKM parameters corresponding to the full class of MFV models. 
With such an approach, the coefficients $P_c$ and $P_r$ become indeed 
model independent and the predictions for $A { \rm (Decay)}$ must depend on 
$v$. Consequently, only certain values for $F_r(v)$, corresponding to 
a particular MFV model, will describe the data correctly
%\end{itemize}

To summarize: 
\begin{itemize}
\item
A model independent analysis of MFV involves {\it eleven} 
parameters:
the real values of the seven master functions $F_r$,
and four CKM parameters, that can be determined independently of $F_r$,
\item
There exist relations between branching ratios that do not involve the 
functions $F_r$ at all. They can be used to test the general concept of MFV 
in a model independent manner.
\item
Explicit calculations indicate that the number of the relevant master 
functions can likely be reduced.
\end{itemize}

These lecture notes provide a rather non-technical description of MFV. 
In Section 2 we discuss briefly the basic theoretical concepts leading 
to the master MFV formula (\ref{mmaster}), recall the CKM matrix and the 
unitarity triangle and 
 discuss the origin of the seven master functions in question. 
We also argue that only five (even only four) functions $F_r$ are 
phenomenologically relevant.
Section 3
is devoted to the determination of the CKM matrix and of the unitarity 
triangle both within the SM and in a general MFV model. 
 In Section 4 we list most interesting MFV relations between various 
observables.
In Section 5 we 
outline  procedures for the determination of the values of the seven (five) 
master functions from the present and forthcoming data. 
In Section 6 we present the results
in a specific MFV model: the SM with one 
universal large extra dimension.  
We give a brief summary in Section 7. 

These lectures are 
complementary to my recent Schladming lectures \cite{Schladming}. The 
discussion of various types of CP violation, the detailed presentation 
of the methods for the determination of the angles of the unitarity triangle 
and detailed treatment of $\epe$ and $K\to\pi\nu\bar\nu$ can be found there, 
in my Erice lectures \cite{Erice} and \cite{BF97}. 
More technical aspects of the field 
are given in my Les Houches lectures \cite{AJBLH}, in the review 
\cite{BBL} and in a very 
recent TASI lectures on effective field theories \cite{IROTH}.
Finally, I would like to recommend
the working group reports
\cite{CERNCKM,BABAR,LHCB,FERMILAB} and most recent reviews \cite{REV}. 
A lot of material, but an exciting one.

\section{Theory of MFV}
\subsection{Preface}
The MFV models have already been investigated in the 1980's and in the first 
half of the 1990's. However, their 
precise formulation has been given only recently, first in \cite{MFV0}  
in the context of the MSSM with minimal flavour violation and subsequently 
in a model independent manner in \cite{UUT}. This particular formulation 
is very simple.
It collects in one class models in which
\begin{itemize}
\item
All flavour changing transitions are governed by the CKM matrix with the 
CKM phase being the only source of CP violation, in particular there are no 
FCNC processes at the tree level,
\item
The only relevant operators in the effective Hamiltonian below the weak scale
are those that are also relevant in the SM.
\end{itemize}
The SM, the Two Higgs Doublet Models I and II, the MSSM with minimal flavour 
violation, all with not too large $\tan\beta$, and the SM with one extra
universal large dimension belong to this class.

Another 
formulation, a profound one, that uses flavour symmetries has been 
presented in \cite{AMGIISST}. 
Similar ideas in the context of specific new physics scenarios can be 
found in \cite{FL1,FL2,FL3}.
While there is a 
considerable overlap of the approach in \cite{AMGIISST} with \cite{UUT} 
there are differences 
between these two 
approaches that are phenomenologically relevant. In short: in this
formulation new operators that are strongly suppressed in the SM are
admitted, modifying in certain cases the phenomenology of weak decays in a 
significant manner. This is in particular the case of the MSSM
with minimal flavour violation but large $\tan\beta$, where scalar operators 
originating from Higgs penguins become important.
Similar comments apply to \cite{BOEWKRUR}. In these lectures we will only 
discuss the MFV models defined, as in \cite{UUT}, by the two items above.

The 
first model independent analysis of the MFV models appeared to my knowledge 
already in \cite{ALIMFV} with the restriction to $B$ decays. 
However, the approach of \cite{ALIMFV}
 differs from 
the one presented here in that the basic phenomenological quantities 
in \cite{ALIMFV}  are the values of Wilson 
coefficients of the relevant operators evaluated at $\mu=\ord(m_b)$  and 
not the values of the master 
functions. The most recent analysis of this type can be found in 
\cite{Ali:2002jg}.

If one assumes in accordance with the experimental findings that all new 
particles have masses larger than $M_{W,Z}$ it is more useful to 
describe 
the MFV models directly in terms of the master functions $F_r(v)$ rather
than with the help of the Wilson coefficients normalized at low energy scales.
We will emphasize this in more detail below.

The formulation of 
decay amplitudes in terms of seven master functions has been proposed for 
the first time in \cite{PBE,BH92} in the context of the SM. The correlations 
between various 
functions and decays within the SM have been first presented in \cite{BH92}. 
Subsequent analyses of this type, still within the SM, can be found in 
\cite{BBSIN,Perez}. 
These lectures discuss the general aspects of MFV models in terms of the
master functions, reviewing the results in the literature and presenting 
new ones.

\subsection{The Basis}
The master formula (\ref{mmaster}), obtained first within the SM a long time
ago \cite{PBE}, is based on the operator product 
expansion (OPE) \cite{OPE} that allows to separate short ($\mu_{SD}$) and 
long 
($\mu_{LD}$) distance 
contributions to weak amplitudes and on the renormalization group (RG) 
methods that allow to sum large logarithms $\log \mu_{SD}/\mu_{LD}$  to 
all orders in perturbation theory. The full exposition of these methods 
can be found in \cite{AJBLH,BBL}. 

The OPE allows to write
the effective weak Hamiltonian simply as follows
\be\label{b1}
{\cal H}_{eff}=\frac{G_F}{\sqrt{2}}\sum_i V^i_{\rm CKM}C_i(\mu)Q_i~.
\ee
Here $G_F$ is the Fermi constant and $Q_i$ are the relevant local
operators which govern the decays in question. 
They are built out of quark and lepton fields.
The CKM
factors $V^i_{\rm CKM}$ \cite{CAB,KM} 
and the Wilson coefficients $C_i(\mu)$ describe the 
strength with which a given operator enters the Hamiltonian.
The Wilson coefficients can be considered as scale dependent
``couplings'' related to ``vertices'' $Q_i$ and 
can be calculated using perturbative methods as long as $\mu$ is
not too small.

An amplitude for a decay of a given meson 
$M= K, B,..$ into a final state $F=\pi\nu\bar\nu,~\pi\pi,~DK$ is then
simply given by
\be\label{amp5}
A(M\to F)=\langle F|{\cal H}_{eff}|M\rangle
=\frac{G_F}{\sqrt{2}}\sum_i V^i_{CKM}C_i(\mu)\langle F|Q_i(\mu)|M\rangle,
\ee
where $\langle F|Q_i(\mu)|M\rangle$ 
are the matrix elements of $Q_i$ between $M$ and $F$, evaluated at the
renormalization scale $\mu$. 

The essential virtue of OPE is that it allows to separate the problem
of calculating the amplitude
$A(M\to F)$ into two distinct parts: the {\it short distance}
(perturbative) calculation of the coefficients $C_i(\mu)$ and 
the {\it long-distance} (generally non-perturbative) calculation of 
the matrix elements $\langle Q_i(\mu)\rangle$. The scale $\mu$
separates, roughly speaking, the physics contributions into short
distance contributions contained in $C_i(\mu)$ and the long distance 
contributions
contained in $\langle Q_i(\mu)\rangle$. 
Thus $C_i$ include the top quark contributions and
contributions from other heavy particles such as W-, Z-bosons, charged
Higgs particles,  supersymmetric particles and Kaluza--Klein modes in models 
with large extra dimensions. 
Consequently, $C_i(\mu)$ depend generally 
on $m_t$ and also on the masses of new particles if extensions of the 
SM are considered. This dependence can be found by evaluating 
so-called {\it box} and {\it penguin} diagrams with full W-, Z-, top- and 
new particle exchanges and {\it properly} including short distance QCD 
effects. The latter govern the $\mu$-dependence of $C_i(\mu)$.

The value of $\mu$ can be chosen arbitrarily but the final result
must be $\mu$-independent.
Therefore 
the $\mu$-dependence of $C_i(\mu)$ has to cancel the 
$\mu$-dependence of $\langle Q_i(\mu)\rangle$. 
The same comments apply to the renormalization scheme dependence of 
$C_i(\mu)$ and $\langle Q_i(\mu)\rangle$.

Now due to the fact that for low energy processes the appropriate scale 
 $\mu$ is much smaller than $M_{W,Z},~ m_t$, large logarithms 
$\ln\mw/\mu$ compensate in the evaluation of
$C_i(\mu)$ the smallness of the QCD coupling constant $\alpha_s$ and 
terms $\alpha^n_s (\ln\mw/\mu)^n$, $\alpha^n_s (\ln\mw/\mu)^{n-1}$ 
etc. have to be resummed to all orders in $\alpha_s$ before a reliable 
result for $C_i$ can be obtained.
This can be done very efficiently by means of the renormalization group
methods. 
The resulting {\it renormalization group improved} perturbative
expansion for $C_i(\mu)$ in terms of the effective coupling constant 
$\alpha_s(\mu)$ does not involve large logarithms and is more reliable.
The related technical issues are discussed in detail in \cite{AJBLH}
and \cite{BBL}. It should be emphasized that by 2003 the next-to-leading 
(NLO) QCD and QED corrections to all relevant weak decay processes 
in the SM and to a large extent in the MSSM are known. But this is another 
story.

Clearly, in order to calculate the amplitude $A(M\to F)$ the matrix 
elements $\langle Q_i(\mu)\rangle$ have to be evaluated. 
Since they involve long distance contributions one is forced in
this case to use non-perturbative methods such as lattice calculations, the
$1/N$ expansion ($N$ is the number of colours), QCD sum rules, hadronic sum rules
and chiral perturbation theory. In the case of $B$ meson decays,
the {\it Heavy Quark Effective Theory} (HQET), {\it Heavy Quark
Expansions} (HQE) and in the case of nonleptonic decays QCD factorization 
(QCDF) and PQCD approach also turn out to be useful tools.
However, all these non-perturbative methods have some limitations.
Consequently the dominant theoretical uncertainties in the decay amplitudes
reside in the matrix elements $\langle Q_i(\mu)\rangle$ and non-perturbative 
parameters present in HQET, HQE, QCDF and PQCD.
These issues are reviewed in \cite{CERNCKM}, where the references to the 
original literature can be found.

The fact that in many cases the matrix elements $\langle Q_i(\mu)\rangle$
 cannot be reliably
calculated at present, is very unfortunate. The main goals of the
experimental studies of weak decays is the determination of the CKM factors 
$V_{\rm CKM}$
and the search for the physics beyond the SM. Without a reliable
estimate of $\langle Q_i(\mu)\rangle$ these goals cannot be achieved unless 
these matrix elements can be determined experimentally or removed from the 
final measurable quantities
by taking suitable ratios and combinations of decay amplitudes or branching
ratios. 
Flavour symmetries like $SU(2)_{\rm F}$ and 
$SU(3)_{\rm F}$ relating various
matrix elements can be useful in this respect, provided flavour
symmetry breaking effects can be reliably calculated.

\subsection{The Basic Idea}
By now all this is standard. It is also standard to choose $\mu$ in
(\ref{amp5}) of $\ord(m_b)$ and $\ord(1-2\gev)$ for $B$ and $K$ decays, 
respectively. But this is certainly {\it not} what we want to do here. 
If we want to expose the short disctance structure of flavour physics and 
in particular the new physics contributions, it is much more useful to 
choose $\mu$ as high as possible but still low enough so that below this 
$\mu$ the physics is fully described by the SM \cite{PBE}. 
We will denote this scale 
by $\mu_0$. This scale is $\ord(M_W,m_t)$.

We are now in a position to
demonstrate that indeed the formula (\ref{amp5}) with a low $\mu$ can 
be cast into the 
master formula (\ref{mmaster}). To this end we first express $C_i(\mu)$ 
in terms of $C_k(\mu_0)$:
\be\label{A1}
C_i(\mu)=\sum_k U_{ik}(\mu,\mu_0) C_k(\mu_0)
\ee
where $U_{ik}(\mu,\mu_0)$ are the elements of the renormalization group 
evolution matrix (from the high scale $\mu_0$ to a low scale $\mu$) 
that depends on the anomalous dimensions of the operators 
$Q_i$ and on the $\beta$ function that governs the evolution of the QCD 
coupling constant. $C_k(\mu_0)$ can be found in the process of the matching
of the full and the effective theory or equvalently integrating out the heavy 
fields with masses larger than $\mu_0$. They are linear combinations of the 
master functions $F_r(v)$ mentioned in the opening section, so that we have
\be\label{A2}
C_k(\mu_0)=g_k+\sum_r h_{kr} F_r(v)
\ee
where $g_k$ and $h_{kr}$ are $v$--independent.
Inserting (\ref{A1}) and (\ref{A2}) into (\ref{amp5}), we easily obtain
(\ref{mmaster}) with $g_k$ and $h_{kr}$ absorbed into $P_c$ and $P_r$, 
respectively. The QCD factors $\eta_i^{  QCD}$ mentioned in  
Section 1 are constructed from $U_{ik}(\mu,\mu_0)$ and are fully calculable 
in the SM. Explicitly we have
\be
P_c({\rm Decay})= \frac{G_F}{\sqrt{2}}\sum_{i,k} V^i_c \langle
Q_i(\mu)\rangle U^c_{ik}(\mu,\mu_0) g_k
\ee
\be
P_r({\rm Decay} ) =\frac{G_F}{\sqrt{2}}\sum_{i,k} V^i_t \langle
Q_i(\mu)\rangle U^t_{ik}(\mu,\mu_0) h_{kr}
\ee
where the additional indices $c$ and $t$ indicate that the CKM parameters 
$V^i_{c,t}$ and QCD corrections in $P_c$ and $P_r$ differ generally from 
each other.
In practice it is convenient to factor out the CKM dependence from 
$P_c$ and $P_r$ and we will do it later on, but at this stage 
it is not necessary. It is more important to realize already here that
\begin{itemize}
\item
$\langle Q_i(\mu)\rangle$ and $U_{ik}(\mu,\mu_0)$ can be calculated fully 
within the SM as functions of $\alpha_s$, $\alpha_{\rm QED}$ and the masses of 
light quarks,
\item
$P_c$ and $P_r$ are $\mu$ independent as the $\mu$ dependence cancels between
$\langle Q_i(\mu)\rangle$ and $U_{ik}(\mu,\mu_0)$,
\item
the coefficients $g_k$ and $h_{kr}$ are process independent (the basic 
property of Wilson coefficients) and can always be chosen so that they
are universal within the MFV models considered here.
\end{itemize} 
This discussion shows that the only ``unknowns" in the master formula 
(\ref{mmaster}) are the CKM parameters, hidden in $P_r$ and $P_c$, and 
the master functions $F_r(v)$. The CKM parameters cannot be 
calculated within the SM and to my knowledge in any MFV model on the 
market. Consequently they have to be extracted from experiment. The 
functions $F_r$, on the other hand, are calculable in a given MFV model 
in perturbation theory as $\alpha_s(\mu_0)$ is small and if necessary in 
a renormalization group improved perturbation theory in the presence of
vastly different scales. But if we want to be fully model independent 
within the class of different MFV models, the values of the functions 
$F_r$ should be directly extracted from experiment.

While the derivation above is given explicitly for exclusive $\Delta F=1$ 
decays, it is
straightforward to generalize it to $\Delta F=2$ transitions and 
inclusive decays. Indeed, the effective
Hamiltonian in (\ref{b1}) is also the fundamental object in inclusive 
decays.

Let us next compare our formulation of FCNC processes with the one in 
\cite{ALIMFV,Ali:2002jg}.
We have emphasized 
here the parametrization of MFV models in terms of master functions, rather 
than Wilson coefficients of certain local operators as done in
\cite{ALIMFV}. 
While the latter formulation involves scales as low as $\ord(m_b)$
 and $\ord(1\gev)$, the 
former one exhibits more transparently the short distance contributions at 
scales $\ord(M_W,m_t)$ and higher. 
The formulation presented here has also the advantage that 
it allows to formulate the $B$ and $K$ decays in terms of the same building 
blocks, the master functions. This allows to study transparently the 
correlations between not only different $B$ or $K$ decays but also between 
$B$ and 
$K$ decays. This clearly is much harder when working directly with the Wilson 
coefficients evaluated at low energy scales. In particular the study of 
the correlations between $K$ and $B$
decays is very difficult as the Wilson coefficients in $K$ and $B$ decays, 
with a few exceptions, involve different renormalization scales. 

Let us finally compare our definition of MFV with the one of 
\cite{AMGIISST}. In this paper the CKM matrix still remains to be the only 
origin of flavour violation but new local operators with new Dirac structures
are admitted to contribute significantly. This is in particular the case of 
MSSM with large $\tan\beta$ in which Higgs penguins, very strongly suppressed
in our version of MFV, contribute in a very important manner (see reviews in 
\cite{CR,Dedes}).
For instance the branching ratios for $B^0_{s,d}\to \mu^+\mu^-$ can be 
enhanced at $\tan\beta=\ord(50)$ 
up to three orders of magnitude with respect to the SM and MFV models 
defined here. I find it difficult to put in the same class models whose 
predictions for certain observables differ by orders of magnitude. 

Moreover, in the framework of \cite{AMGIISST},
 the new physics contributions cannot 
be always taken into account by simply modifying the master functions as 
done in our approach.
As a result this formulation  involves more independent 
parameters and
some very useful relations discussed in Section 4 that are valid in MFV 
models discussed here can be violated in the class of models considered in 
\cite{AMGIISST} even for a low  $\tan\beta$ and in models with a single Higgs 
doublet.
In particular the correlations between semileptonic and nonleptonic decays 
present in our approach are essentially absent there.
All these relations and correlations are important phenomenologically because
their violation would immediately signal the presence of new phases and/or
new local operators that are irrelevant in the SM. 
Only time will show which of these two 
frameworks is closer to the data.

These comments should not be considered by any means 
as a critique of the formulation in \cite{AMGIISST}, that I find very
elegant,   but as the first step 
beyond the SM, the definition of MFV given in \cite{UUT} and used in these
lectures appears  more useful to me. On the other hand the authors in
\cite{AMGIISST}, similarly to \cite{PBE,BH92} and us here, use as free 
parameters quantities normalized at a high scale $\mu_0$ so that in their 
formulation certain correlations between $K$ and $B$ decays can also be
transparently seen.

After this general discussion let us have a closer look at the CKM matrix
and subsequently the functions $F_r(v)$.

\subsection{CKM Matrix and the Unitarity Triangle (UT)}
The unitary CKM matrix \cite{CAB,KM} connects  the {\it weak
eigenstates} $(d^\prime,s^\prime,b^\prime)$ and 
 the corresponding {\it mass eigenstates} $d,s,b$:
\begin{equation}\label{2.67}
\left(\begin{array}{c}
d^\prime \\ s^\prime \\ b^\prime
\end{array}\right)=
\left(\begin{array}{ccc}
V_{ud}&V_{us}&V_{ub}\\
V_{cd}&V_{cs}&V_{cb}\\
V_{td}&V_{ts}&V_{tb}
\end{array}\right)
\left(\begin{array}{c}
d \\ s \\ b
\end{array}\right)\equiv\hat V_{\rm CKM}\left(\begin{array}{c}
d \\ s \\ b
\end{array}\right).
\end{equation}

Many parametrizations of the CKM
matrix have been proposed in the literature. The classification of different 
parametrizations can be found in \cite{FX1}. While the so called 
standard parametrization \cite{CHAU} 
\begin{equation}\label{2.72}
\hat V_{\rm CKM}=
\left(\begin{array}{ccc}
c_{12}c_{13}&s_{12}c_{13}&s_{13}e^{-i\delta}\\ -s_{12}c_{23}
-c_{12}s_{23}s_{13}e^{i\delta}&c_{12}c_{23}-s_{12}s_{23}s_{13}e^{i\delta}&
s_{23}c_{13}\\ s_{12}s_{23}-c_{12}c_{23}s_{13}e^{i\delta}&-s_{23}c_{12}
-s_{12}c_{23}s_{13}e^{i\delta}&c_{23}c_{13}
\end{array}\right)\,,
\end{equation}
with
$c_{ij}=\cos\theta_{ij}$ and $s_{ij}=\sin\theta_{ij}$ 
($i,j=1,2,3$) and the complex phase $\delta$ necessary for {\rm CP} violation,
should be recommended \cite{PDG} 
for any numerical 
analysis, a generalization of the Wolfenstein parametrization \cite{WO} as 
presented in \cite{BLO} is more suitable for these lectures. 
On the one hand it is more transparent than the standard parametrization and 
on the other hand it  satisfies the unitarity 
of the CKM matrix to higher accuracy  than the original parametrization 
in \cite{WO}. 

To this end we make the following change of variables in
the standard parametrization (\ref{2.72}) 
\cite{BLO,schubert}
\begin{equation}\label{2.77} 
s_{12}=\lambda\,,
\qquad
s_{23}=A \lambda^2\,,
\qquad
s_{13} e^{-i\delta}=A \lambda^3 (\varrho-i \eta)
\end{equation}
where
\begin{equation}\label{2.76}
\lambda, \qquad A, \qquad \varrho, \qquad \eta \, 
\end{equation}
are the Wolfenstein parameters
with $\lambda\approx 0.22$ being an expansion parameter. We find then 
\be\label{f1}
V_{ud}=1-\frac{1}{2}\lambda^2-\frac{1}{8}\lambda^4, \qquad
V_{cs}= 1-\frac{1}{2}\lambda^2-\frac{1}{8}\lambda^4(1+4 A^2),
\ee
\be
V_{tb}=1-\frac{1}{2} A^2\lambda^4, \qquad
V_{cd}=-\lambda+\frac{1}{2} A^2\lambda^5 [1-2 (\varrho+i \eta)],
\ee
\be\label{VUS}
V_{us}=\lambda+\ord(\lambda^7),\qquad 
V_{ub}=A \lambda^3 (\varrho-i \eta), \qquad 
V_{cb}=A\lambda^2+\ord(\lambda^8),
\ee
\begin{equation}\label{2.83d}
 V_{ts}= -A\lambda^2+\frac{1}{2}A\lambda^4[1-2 (\varrho+i\eta)],
\qquad V_{td}=A\lambda^3(1-\bar\varrho-i\bar\eta)
\end{equation}
where terms 
$\ord(\lambda^6)$ and higher order terms have been neglected.
A non-vanishing $\eta$ is responsible for CP violation in the MFV models.
 It plays 
the role of $\delta$ in the standard parametrization.
Finally, the barred variables in (\ref{2.83d}) are given by
\cite{BLO}
\begin{equation}\label{2.88d}
\bar\varrho=\varrho (1-\frac{\lambda^2}{2}),
\qquad
\bar\eta=\eta (1-\frac{\lambda^2}{2}).
\end{equation}

Now, the unitarity of the CKM-matrix implies various relations between its
elements. In particular, we have
\begin{equation}\label{2.87h}
V_{ud}^{}V_{ub}^* + V_{cd}^{}V_{cb}^* + V_{td}^{}V_{tb}^* =0.
\end{equation}
The relation (\ref{2.87h})  can be
represented as a ``unitarity'' triangle in the complex 
$(\bar\varrho,\bar\eta)$ plane. 
One can construct five additional  unitarity triangles \cite{Kayser} 
corresponding to other unitarity relations.

Noting that to an excellent accuracy $V_{cd}^{}V_{cb}^*$ is real with
$| V_{cd}^{}V_{cb}^*|=A\lambda^3+\ord(\lambda^7)$ and
rescaling all terms in (\ref{2.87h}) by $A \lambda^3$ 
we indeed find that the relation (\ref{2.87h}) can be represented 
as a triangle 
in the complex $(\bar\varrho,\bar\eta)$ plane 
as shown in fig.~\ref{fig:utriangle}. Let us collect useful formulae related 
to this triangle:

\begin{figure}[hbt]
\vspace{0.10in}
\centerline{
\epsfysize=1.7in
\epsffile{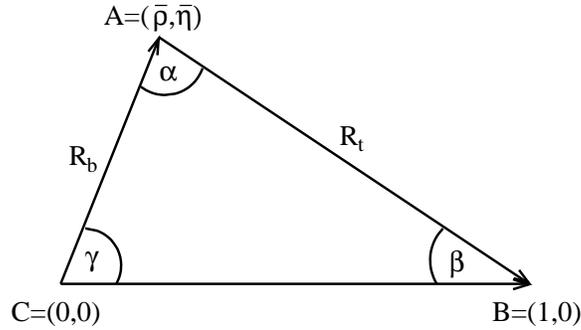}
}
\vspace{0.08in}
\caption{Unitarity Triangle.}\label{fig:utriangle}
\end{figure}

\bi
\item
We can express $\sin(2\beta$)
in terms of $(\bar\varrho,\bar\eta)$:
\begin{equation}\label{2.90}
\sin(2\beta)=\frac{2\bar\eta(1-\bar\varrho)}{(1-\bar\varrho)^2 + \bar\eta^2}.
\end{equation}
\item
The lengths $CA$ and $BA$ are given respectively by \cite{BH92,BLO}
\begin{equation}\label{2.94}
R_b \equiv \frac{| V_{ud}^{}V^*_{ub}|}{| V_{cd}^{}V^*_{cb}|}
= \sqrt{\bar\varrho^2 +\bar\eta^2}
= (1-\frac{\lambda^2}{2})\frac{1}{\lambda}
\left| \frac{V_{ub}}{V_{cb}} \right|,
\end{equation}
\begin{equation}\label{2.95}
R_t \equiv \frac{| V_{td}^{}V^*_{tb}|}{| V_{cd}^{}V^*_{cb}|} =
 \sqrt{(1-\bar\varrho)^2 +\bar\eta^2}
=\frac{1}{\lambda} \left| \frac{V_{td}}{V_{cb}} \right|.
\end{equation}
\item
The angles $\beta$ and $\gamma=\delta$ of the unitarity triangle 
are related
directly to the complex phases of the CKM elements $V_{td}$ and
$V_{ub}$, respectively, through
\beq\label{e417}
V_{td}=|V_{td}|e^{-i\beta},\quad V_{ub}=|V_{ub}|e^{-i\gamma}.
\eeq
\item
The unitarity relation (\ref{2.87h}) can be rewritten as
\be\label{RbRt}
R_b e^{i\gamma} +R_t e^{-i\beta}=1~.
\ee
\item
The angle $\alpha$ can be obtained through the relation
\beq\label{e419}
\alpha+\beta+\gamma=180^\circ~.
\eeq
\ei

Formula (\ref{RbRt}) shows transparently that the knowledge of
$(R_t,\beta)$ allows to determine $(R_b,\gamma)$ through 
\be\label{VUBG}
R_b=\sqrt{1+R_t^2-2 R_t\cos\beta},\qquad
\cot\gamma=\frac{1-R_t\cos\beta}{R_t\sin\beta}.
\ee
Similarly, $(R_t,\beta)$ can be expressed through $(R_b,\gamma)$:
\be\label{VTDG}
R_t=\sqrt{1+R_b^2-2 R_b\cos\gamma},\qquad
\cot\beta=\frac{1-R_b\cos\gamma}{R_b\sin\gamma}.
\ee
These relations are remarkable. They imply that the knowledge 
of the coupling $V_{td}$ between $t$ and $d$ quarks allows to deduce the 
strength of the corresponding coupling $V_{ub}$ between $u$ and $b$ quarks 
and vice versa.

The triangle depicted in fig. \ref{fig:utriangle}, $|V_{us}|$ 
and $\vcb$ give the full description of the CKM matrix. 
Looking at the expressions for $R_b$ and $R_t$, we observe that within
the MFV models the measurements of four CP
{\it conserving } decays sensitive to $|V_{us}|$, $|V_{ub}|$,   
$|V_{cb}|$ and $|V_{td}|$ can tell us whether CP violation
($\bar\eta \not= 0$ or $\gamma \not=0,\pi$) is predicted in the MFV models. 
This fact is often used to determine
the angles of the unitarity triangle without the study of CP-violating
quantities.

\subsection{Master Functions}
\begin{figure}[hbt]
\centerline{
\epsfysize=4.3in
\epsffile{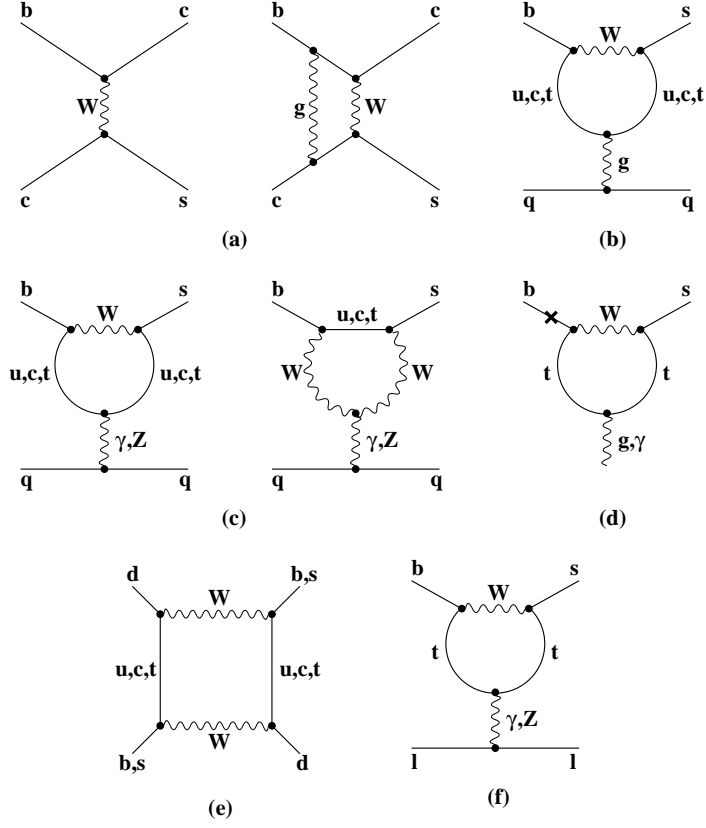}
}
\caption{Typical Penguin and Box Diagrams in the SM.}
\label{fig:fdia}
\end{figure}

The master functions $F_r(v)$ originate from
various penguin and box diagrams. Some examples relevant for the SM
are shown in fig.~\ref{fig:fdia}. Analogous diagrams are present in 
the extensions of the SM. 

In order to find the master functions $F_r(v)$, we first express
the penguin vertices (including electroweak counter terms)
in terms of the functions
$C$ ($Z^0$ penguin), $D$ ($\gamma$ penguin), $E$ (gluon penguin), 
$D'$ ($\gamma$-magnetic penguin) and $E'$ (chromomagnetic 
penguin). In the  
't Hooft--Feynman gauge for the $W^\pm$  propagator they are  
given as follows:
\begin{equation}\label{ZRULE}
 \bar s Z d~ =~i \lambda_t {{G_{\rm F}}\over{\sqrt 2}} {{g_2}\over{2\pi^2}} 
   {{M^2_W}\over{\cos\theta_{w}}} C(v) \bar s \gamma_\mu 
   (1-\gamma_5)d
\end{equation}
\begin{equation}\label{Ddef}
 \bar s\gamma d~ =~- i\lambda_t {{G_{\rm F}}\over{\sqrt 2}} {{e}\over{8\pi^2}}
   D(v) \bar s (q^2\gamma_\mu - q_\mu \not\!q)(1-\gamma_5)d 
\end{equation}
\begin{equation}\label{Edef}
 \bar s G^a d~ =~ -i\lambda_t{{G_{\rm F}}\over{\sqrt 
2}} {{g_s}\over{8\pi^2}}
   E(v) \bar s_{\alpha}(q^2\gamma_\mu - q_\mu \not\!q)
(1-\gamma_5)T^a_{\alpha\beta}d_\beta 
\end{equation}
\begin{equation}\label{MGP}
 \bar s \gamma' b~ =~i\bar\lambda_t {{G_{\rm F}}\over{\sqrt 2}} {{e}\over
   {8\pi^2}} D'(v) \bar s \lbrack i\sigma_{\mu\lambda} q^\lambda
   \lbrack m_b (1+\gamma_5) \rbrack\rbrack b
\end{equation}
\begin{equation}\label{FRF}
 \bar s G'^a b~ =~ 
i\bar\lambda_t{{G_{\rm F}}\over{\sqrt 2}}{{g_s}\over{8\pi^2}}
   E'(v)\bar s_{\alpha} \lbrack i\sigma_{\mu\lambda} q^\lambda
   \lbrack m_b (1+\gamma_5) \rbrack\rbrack T^a_{\alpha\beta} b_\beta \,,
\end{equation}
where $G_F$ is the Fermi constant, $\theta_w$ is the weak mixing angle and 
\be
\lambda_t=V^*_{ts}V_{td},\quad\quad \bar\lambda_t=V^*_{ts}V_{tb}~. 
\ee
In these vertices
$q_\mu$ is the {\it outgoing} gluon or photon momentum
and $T^a$ are colour matrices.
The last two vertices involve an on-shell photon
and an on-shell gluon, respectively.
We have set $m_s=0$ in these vertices. 

Similarly we can define the box function $S$ ($\Delta F=2$ transitions), 
as well as $\Delta F=1$ box functions
$B^{\nu\bar\nu}$ and $B^{\mu\bar\mu}$ relevant for decays with 
${\nu\bar\nu}$  and ${\mu\bar\mu}$ in the final state, respectively.
Explicitly:
\begin{equation}\label{FR}
  {\rm Box} (\Delta S = 2)
~=~ \lambda^2_i {{G^2_{\rm F}}\over{16\pi^2}} \mw^2S(v) 
   (\bar s d)_{V-A} (\bar s d)_{V-A} 
\end{equation}
\begin{equation}
 {\rm Box}(T_3= 1/2)~ =~ \lambda_i {{G_{\rm F}}\over{\sqrt 2}}
{{\alpha}\over{2\pi \sin^2\theta_{\rm w}}} B^{\nu\bar\nu}(v)(\bar s d)_{V-A} 
(\bar\nu\nu)_{V-A}
\end{equation}
\begin{equation}
{\rm Box}(T_3= -1/2)~ =-~ \lambda_i {{G_{\rm F}}\over{\sqrt 2}}
   {{\alpha}\over{2\pi \sin^2\theta_{\rm w}}} B^{\mu\bar\mu}(v)
 (\bar s d)_{V-A} 
   (\bar\mu\mu)_{V-A}
\end{equation}
where $T_3$ is the weak isospin of the final lepton.
In the case of $\Delta F=1$ box diagrams with $u\bar u$ and $d\bar d$ in
the final state we have to an excellent approximation
\be\label{NON}
B^{u\bar u}(v)=B^{\nu\bar\nu}(v), \qquad B^{d\bar d}(v)=B^{\mu\bar\mu}(v),
\ee
simply because these relations are perfect in the SM and the contributions 
of new physics to $\Delta F=1$ box diagrams turn out to be very small 
in the MFV models. In case of questions related to ``i" factors and signs 
in the formulae above, the interested reader is asked to consult 
\cite{Erice,AJBLH}.

While the $\Delta F=2$ box function $S$ and the penguin functions 
$E$, $D'$ and $E'$ are gauge independent, this is not the case for 
$C$, $D$ and the $\Delta F=1$ box diagram functions 
$B^{\nu\bar\nu}$ and $B^{\mu\bar\mu}$.
In the phenomenological applications it is more 
convenient to work with gauge independent functions \cite{PBE}
\begin{equation}\label{XYZ} 
X(v)=C(v)+B^{\nu\bar\nu}(v),\qquad  
Y(v)  =C(v)+B^{\mu\bar\mu}(v), \qquad
Z(v)  =C(v)+\frac{1}{4}D(v).
\end{equation}
Indeed, the box diagrams have the Dirac structure $(V-A)\otimes (V-A)$, 
the $Z^0$ penguin diagram has the $(V-A)\otimes(V-A)$ and 
$(V-A)\otimes V$ components and the $\gamma$ penguin is pure
$(V-A)\otimes V$.
The $X$ and $Y$ correspond then to linear combinations of the 
$(V-A)\otimes(V-A)$ component 
of the $Z^0$ penguin diagram and box diagrams with final quarks and leptons 
having weak isospin $T_3=1/2$ and $T_3=-1/2$, respectively. $Z$ corresponds 
to the linear combination of  the $(V-A)\otimes V$ component 
of the $Z^0$ penguin diagram and the $\gamma$ penguin.

Then the set of seven gauge independent master functions which govern
the FCNC processes in the MFV models is given by:
\be\label{masterf}
S(v),~X(v),~Y(v),~Z(v),~E(v),~ D'(v),~ E'(v)~.
\ee

In the SM we have  to a very good
approximation ($x_t=m^2_t/\mw^2$):
\begin{equation}\label{S0}
 S_0(x_t)=2.40~\left(\frac{\mt}{167\gev}\right)^{1.52},
\ee
\be\label{XA0}
X_0(x_t)=1.53~\left(\frac{\mt}{167\gev}\right)^{1.15},
\quad\quad
Y_0(x_t)=0.98~\left(\frac{\mt}{167\gev}\right)^{1.56},
\end{equation}
\begin{equation}
 Z_0(x_t)=0.68~\left(\frac{\mt}{167\gev}\right)^{1.86},\quad\quad
   E_0(x_t)= 0.27~\left(\frac{\mt}{167\gev}\right)^{-1.02},
\end{equation}
\begin{equation}
 D'_0(x_t)=0.38~\left(\frac{\mt}{167\gev}\right)^{0.60}, \quad\quad 
E'_0(x_t)=0.19~\left(\frac{\mt}{167\gev}\right)^{0.38}.
\end{equation}
The subscript ``$0$'' indicates that 
these functions
do not include QCD corrections to the relevant penguin and box diagrams.
Exact expressions for all functions can be found in \cite{AJBLH}.
Let us also recall that in the SM
\be\label{B0}
B^{\nu\bar\nu}(v)=-4~ B_0(x_t),\qquad  
B^{\mu\bar\mu}(v)=-B_0(x_t)
\end{equation}
with $B_0(x_t)=-0.182$ for $m_t=167\gev$.

Generally, several master functions contribute to a given decay,
although decays exist which depend only on a single function.
We have the following correspondence between the most interesting FCNC
processes and the master functions in the MFV models \cite{BH92}:
\begin{center}
\begin{tabular}{lcl}
$K^0-\bar K^0$-mixing ($\varepsilon_K$) 
&\qquad\qquad& $S(v)$ \\
$B_{d,s}^0-\bar B_{d,s}^0$-mixing ($\Delta M_{s,d}$) 
&\qquad\qquad& $S(v)$ \\
$K \to \pi \nu \bar\nu$, $B \to X_{d,s} \nu \bar\nu$ 
&\qquad\qquad& $X(v)$ \\
$K_{\rm L}\to \mu \bar\mu$, $B_{d,s} \to l\bar l$ &\qquad\qquad& $Y(v)$ \\
$K_{\rm L} \to \pi^0 e^+ e^-$ &\qquad\qquad& $Y(v)$, $Z(v)$, 
$E(v)$ \\
$\varepsilon'$, Nonleptonic $\Delta B=1$, $\Delta S=1$ &\qquad\qquad& $X(v)$,
$Y(v)$, $Z(v)$,
$E(v)$ \\
$B \to X_s \gamma$ &\qquad\qquad& $D'(v)$, $E'(v)$ \\
$B \to X_s~{\rm gluon}$ &\qquad\qquad& $E'(v)$ \\
$B \to X_s l^+ l^-$ &\qquad\qquad&
$Y(v)$, $Z(v)$, $E(v)$, $D'(v)$, $E'(v)$
\end{tabular}
\end{center}

This table means that the observables like branching ratios, mass differences
$\Delta M_{d,s}$ in $B_{d,s}^0-\bar B_{d,s}^0$-mixing and the CP violation 
parameters $\varepsilon$ und $\varepsilon'$, all can be to a very good 
approximation (see below) entirely expressed in
terms of the corresponding master functions and the relevant CKM factors. The
remaining entries in the relevant formulae for these observables are low 
energy parameters present in $P_c$ and $P_r$ that can be calculated within 
the SM.
\subsection{A Guide to the Literature}
The formulae for the processes listed above in the SM, given in terms of 
the master 
functions and CKM factors can be found in many papers. We will recall the
main  structure of these formulae in Section 4. The full list using 
the same notation is given in \cite{BBL}. An update of these formulae 
with additional references is given in two papers on universal extra
dimensions \cite{BSW02,BPSW}, where one has to replace $F_r(x_t,1/R)$ by   
$F_r(v)$ to obtain the formulae in a general MFV model.
The supersymmetric contributions to the functions $S$, $X$,
$Y$, $Z$ and $E$ within the  MSSM with minimal
flavour violation are compiled
in \cite{BRMSSM}. See also \cite{MFV0,ALIMFV,BERTOL,MW96}, where the remaining 
functions can be found. The QCD corrections to these functions can be found 
in \cite{BJW90,BB,BB98,MU98,BGH,QCDMSSM,Bobeth,BSGAMMA}. 
The full set of $F_r(v)$ in the SM with one extra universal dimension
is given in \cite{BSW02,BPSW}.
\subsection{Comments on QCD Corrections}
Let us next clarify the issue of QCD corrections in this formulation. 
To this end let us consider $\Delta M_d$  that parametrizes 
the  $B^0_d-\bar B^0_d$ mixing.
Only one 
master function, $S(v)$, contributes to $\Delta M_d$ in the models considered
here. 
Within the SM we can write
\be\label{DEMd}
\Delta M_d=a\langle Q(\mu_b)\rangle U(\mu_b,\mu_0) C_Q(\mu_0)
\ee
where $\mu_b=\ord(m_b)$ and the coefficient $a$ includes $G^2_F$, the relevant 
CKM factor and some known numerical constants. 
$\langle Q(\mu_b)\rangle$ is the 
matrix element of the relevant local operator 
and $U(\mu_b,\mu_0)$ the corresponding renormalization group factor. 
Calculating the known box diagrams with $W$ and top quark exchanges and 
including $\ord(\alpha_s)$ corrections one finds \cite{BJW90}
\be
C_Q(\mu_0)=S_0(x_t)\left[1+\alpha_s(\mu_0)\Delta(\mu_0)\right].
\ee
As discussed in detail in \cite{AJBLH,BBL,BJW90} and in Section 8 of
\cite{PBE}, the correction $\Delta(\mu_0)$ contains, in addition to a 
complicated $m_t$ dependence, three terms that separately
cancel the operator renormalization scheme dependence of 
$U(\mu_b,\mu_0)$ at the upper end of this renormalization group evolution, 
the $\mu_0$ dependence of $U(\mu_b,\mu_0)$ and the dependence on the scale 
$\mu_t$ in $m_t(\mu_t)$ that could in principle differ from $\mu_0$. Here, in
order to simplify the presentation we will not discuss these three
corrections separately and will proceed as follows.

Noticing \cite{BJW90} 
 that for $\mu_0=\mu_t=m_t$, the correction factor $\Delta(\mu_0)$ is 
essentially independent of $m_t$, it is convenient, in the spirit of our
master formula (\ref{mmaster}), to rewrite (\ref{DEMd}) as follows
\be\label{PS}
\Delta M_d= P_S S_0(x_t), \qquad
P_S=a\langle Q(\mu_b)\rangle U(\mu_b,\mu_0)
\left[1+\alpha_s(\mu_0)\Delta(\mu_0)\right]~.
\ee
The internal charm contributions to $\Delta M_d$ are negiligible and 
$P_c\approx 0$ in this case. We should keep in mind that in writing
(\ref{PS}) we have chosen a special definition of $P_r$ and generally 
only the product $P_r F_r$ is independent of this definition.

Beyond the SM we can generalize (\ref{PS}) to 
\be
\Delta M_d= P_S S(v)
\ee
with $S(v)$ found by calculating the relevant diagrams contributing to 
$\Delta M_d$ in a given MFV extension of the SM. If in this model
\be
C_Q(\mu_0)=\tilde S_0(v)\left[1+\alpha_s(\mu_0)\tilde\Delta(\mu_0)\right],
\ee
then
\be
S(v)=\tilde S_0(v)
\left[1+\alpha_s(\mu_0)(\tilde\Delta(\mu_0)-\Delta(\mu_0))\right]
\ee
with $\tilde S_0(v)$ obtained from the relevant diagrams in this model
without the QCD corrections. In a model independent analysis this discussion 
is unnecessary but if one wants to compare the determined $S(v)$
with the result obtained in a given model, the difference between the 
QCD corrections in the SM and in the considered extension at scales
$\ord(\mu_0)$ has to be taken 
into account as outlined above.
There is no difference between these corrections at lower energy scales.

The second issue is the breakdown of the universality of the master functions
by QCD corrections. Let us consider the $Z^0$ penguin diagrams with 
$q\bar q$ and $l\bar l$ coupled to the lower end of the $Z^0$ propagator 
in fig.~\ref{fig:fdia}c and fig.~\ref{fig:fdia}f, where in the case of $l\bar
l$ only the first diagram has been shown.
 These diagrams 
contribute to nonleptonic  and semileptonic 
 decays, respectively. The one-loop $\bar b Z^0 s $ vertex in diagrams  
with top quark exchanges and other heavy particles is the same in both cases 
even after 
the inclusion of QCD corrections and consequently 
the inclusion of these corrections does not spoil the universality of 
$C(v)$. However, the inclusion of all QCD corrections to the $Z^0$ penguin 
diagrams breaks the universality in question because in nonleptonic decays
there are diagrams with gluons connecting the one-loop $\bar b Z^0 s$ 
vertex with the 
$q\bar q$ line that are clearly absent in the semileptonic case. Similar 
comments apply to $\Delta F=1$ box diagrams with 
$l\bar l$ and $q\bar q$ on the r.h.s of the box diagram.

In $\Delta F=1$ box diagrams the breaking of universality can also take place
in principle even in the absence of QCD corrections because the internal
fermion propagators on the r.h.s of $\Delta F=1$ box diagrams in nonleptonic
decays differ from the ones in semileptonic decays.
These diagrams can be obtained from the diagram e) in fig.~\ref{fig:fdia}.

The studies of these universality breaking corrections in the SM 
\cite{BB,BGH}, 
MSSM \cite{BRMSSM,QCDMSSM} and 
the SM with one universal extra dimension \cite{BSW02,BPSW} 
show that these corrections are very small. In particular, they are 
substantially smaller
than the universal $\ord(\alpha_s)$ corrections to the one loop $Z^0$ vertex.
We expect that this is also the case in other MFV models and
we will assume it in what follows. In the future when the accuracy of data
improves, one could consider the inclusion of these corrections into our 
formalism. While this is straightforward, we think it is an unnecessary 
complication at present. 

The third issue is the correspondence between the decays and the master 
functions given above. In more complicated decays, in which the mixing 
between various operators takes place, it can happen that at NLO and higher 
orders additional functions not listed above could contribute. But they are 
generally suppressed by $\alpha(\mu_0)$ and constitute only small 
corrections. Moreover they can be absorbed in most
cases into the seven master functions.
\subsection{A Useful Conjecture: Reduced Set of Master Functions}
We know from the study of FCNC processes that not all master functions are 
important in a given decay. In particular the contributions of 
the function $E(v)$ to all
semileptonic decays are negligible as it is always multiplied by a small
coefficient $P_E$. It is slightly more important in 
$\varepsilon'/\varepsilon$ but also here it can be neglected to first 
approximation unless one expects order of magnitude enhancements of this 
function by new 
physics contributions. The dominant contribution from 
the gluon penguin to $\varepsilon'/\varepsilon$ comes from the relevant 
$P_c$ term.

While the $\Delta F =1$ box diagram contributions 
$B^{\nu\bar\nu}$ and $B^{\mu\bar\mu}$
are relevant in the SM, 
in all MFV models I know, the corresponding new physics contributions 
to these functions  have been found to be very small. Consequently 
these functions are given to an excellent approximation by (\ref{NON}) and 
 (\ref{B0}).
Similar comments apply to the $D$ function for which the SM value is 
$D_0=-0.48$. 
This means that the new physics contributions to the functions 
$X$, $Y$ and $Z$ enter dominantly through 
the function $C$.   
Now, the function $C$ depends on the 
gauge of the $W$ propagator, but this dependence enters only in the 
subleading terms in $m_t^2/M_W^2$ and is cancelled by the one of the 
box diagrams and photon penguin diagrams.
Plausible general arguments for the  dominance of the $C$ function and 
small new physics effects in the $\Delta F=1$ box diagrams have been given in 
\cite{Buras:1999da,FLISMA}.

Finally, as we will see below, the contributions of $E'(v)$ to 
$B\to X_s\gamma$ and 
$B\to X_s l^+l^-$ are strongly suppressed by small values of $P_{E'}$ and 
to first approximation one can set $E'(v)$ to its SM value in these 
decays.

On the basis of this discussion we conjecture that the set of the independent
master functions in (\ref{masterf}) can be reduced to five, possibly four, 
functions
\be\label{rmasterf}
S(v),~C(v),~(Z(v)),~ D'(v),~ E'(v)~.
\ee
if one does not aim at a high precision. 
In this case the table given in Section 2.5 can be simplified 
significantly :

\begin{center}
\begin{tabular}{lcl}
$K^0-\bar K^0$-mixing ($\varepsilon_K$) 
&\qquad\qquad& $S(v)$ \\
$B_{d,s}^0-\bar B_{d,s}^0$-mixing ($\Delta M_{s,d}$) 
&\qquad\qquad& $S(v)$ \\
$K \to \pi \nu \bar\nu$, $B \to X_{d,s} \nu \bar\nu$ 
&\qquad\qquad& $C(v)$ \\
$K_{\rm L}\to \mu \bar\mu$, $B \to l\bar l$ &\qquad\qquad& $C(v)$ \\
$K_{\rm L} \to \pi^0 e^+ e^-$ &\qquad\qquad& $C(v)$, $(Z(v))$ \\
$\varepsilon'$, Nonleptonic $\Delta B=1$, $\Delta S=1$ 
&\qquad\qquad& $C(v)$, $(Z(v))$ \\
$B \to X_s \gamma$ &\qquad\qquad& $D'(v)$ \\
$B \to X_s~{\rm gluon}$ &\qquad\qquad& $E'(v)$ \\
$B \to X_s l^+ l^-$ &\qquad\qquad&
$C(v)$, $D'(v)$, $(Z(v))$
\end{tabular}
\end{center}
where it is understood that $\Delta F=1$ box functions, the functions $D$ and
$E$ and $E'$ in $B\to X_s\gamma$,
$B\to X_s l^+l^-$ and $\epe$ are set to their SM values. 
In parentheses we give also the full contributions of the $Z(v)$ function 
in case the new physics contributions to $D(v)$ turned out to be more
important than anticipated above.

\subsection{Summary}
We have presented the general structure of the MFV models. In addition to 
the nonperturbative parameters $B_i$, that can be calculated in QCD,
 the basic ingredients in our
formulation are the CKM parameters and the master functions $F_r$. 
We will now proceed to discuss the determination of the CKM parameters.
The procedures for the model independent determination of 
the master functions will be outlined in Section 5.

\section{Determination of the CKM Parameters}\label{UT-Det}
\subsection{The Determination of \boldmath{$|V_{us}|$}, \boldmath{$|V_{ub}|$}
and \boldmath{$|V_{cb}|$}}
These elements are determined in tree level semileptonic $K$ and $B$ decays.
The present situation can be summarized by \cite{CERNCKM} 
\begin{equation}\label{vcb}
|V_{us}| = \lambda =  0.2240 \pm 0.0036\,,
\quad\quad
\vcb=(41.5\pm0.8)\cdot 10^{-3},
\end{equation}
\begin{equation}\label{v13}
\frac{|V_{ub}|}{\vcb}=0.086\pm0.008, \quad\quad
|V_{ub}|=(3.57\pm0.31)\cdot 10^{-3}
\end{equation}
implying
\be\label{ARb}
R_b=0.37\pm 0.04~.
\ee
There is an impressive work done by theorists and experimentalists hidden
behind these numbers. We refer to \cite{CERNCKM} for details.
See also \cite{PDG}.

The information given above determines only the length $R_b$ of the side AC
in the UT. 
While this information appears at first sight to be rather limited, 
it is very important for the following reason. As $|V_{us}|$, $\vcb$, 
 $|V_{ub}|$ and consequently $R_b$ are determined here from tree level 
decays, their
values given above are to an excellent accuracy independent of any 
new physics contributions. They are universal fundamental 
constants valid in any extension of the SM. Therefore their precise 
determinations are of utmost importance. 

\subsection{Completing the Determination of the CKM Matrix}
We have thus determined three out of four parameters of the CKM matrix.
The special feature of this determination was its independence of new 
physics contributions. There are many ways to determine the fourth 
parameter and in particular to construct the UT.
Most promising in this respect are the FCNC processes, both 
CP-violating and CP-conserving. 
These decays are sensitive to the angles $\beta$ and $\gamma$ as well as 
to the length $R_t$ and measuring only one of these three quantities allows to 
find the unitarity triangle provided the universal $R_b$ is known.

Now, we can ask which is the ``best" set
of four parameters in the CKM matrix. 
The most popular at present are the Wolfenstein parameters but in the future 
other sets could turn out to be more useful \cite{BUPAST}. Let us then
briefly discuss this issue. 
I think there is no doubt that $\vus$ and $\vcb$ have to belong to this set. 
Also $R_b$ could in principle be put on this list because of its 
independence of new physics contributions. But the measurement of $\vub$ 
and consequently of $R_b$ is still subject to significant experimental 
and theoretical uncertainties and for the time being I do not think 
it should be put on this list.

A much better {\it third} candidate is the angle $\beta$ in the unitarity 
triangle, the phase of $V_{td}$. It has been measured essentially without
any hadronic uncertainty through the mixing induced CP asymmetry in 
$B_d\to \psi K_S$ \cite{BaBar,Belle} 
and moreover within the MFV models this measurement 
is independent of any new physics contributions. The master functions $F_r$
do not enter the expression for this asymmetry.

The best candidates for the {\it fourth} parameter are the absolute value 
of $\vtd$ or $R_t$, the angle $\gamma$ and the height 
$\bar\eta$ of the UT. These three are generally sensitive to new 
physics contributions but in the MFV models there are ways to
extract $R_t$ and $\gamma$ and consequently also $\bar\eta$ 
independently of new physics contributions. The angle $\gamma$ 
can be extracted from strategies involving charged tree level $B$ decays
that are insensitive to possible new physics effects in the 
particle-antiparticle mixing but this extraction will only be available 
in the second half of this decade. Other strategies for $\gamma$ are 
discussed in \cite{Schladming,REV}. Certainly the popular $B\to K\pi$ 
decays cannot 
be used here as the determination of $\gamma$ in these decays 
depends sensitively on the size of electroweak penguins \cite{BFRS}, even in 
MFV models, and moreover the hadronic uncertainties are substantial.

It appears then that for the near future the fourth useful parameter
is $R_t$ which within the MFV models could soon be determined from 
the ratio $\Delta M_d/\Delta M_s$ ($B^0_{d,s}-\bar B^0_{d,s}$ mixing) 
independently of the size of the master function $S(v)$. The corresponding 
expression will be given below. 

To summarize, the best set of four paramaters of the CKM matrix appears 
at present to be \cite{BUPAST}
\be\label{best}
\vus, \qquad \vcb, \qquad R_t, \qquad \beta
\ee
with $R_t$ determined from $\Delta M_d/\Delta M_s$.
The elements of the CKM matrix are then given as follows 
\cite{BUPAST}:
\be
V_{ud}=1-\frac{1}{2}\lambda^2-\frac{1}{8}\lambda^4 +\ord(\lambda^6),
\qquad
V_{ub}=\frac{\lambda}{1-\lambda^2/2}\vcb \left[1-R_t e^{i\beta}\right],
\ee
\be
V_{cd}=-\lambda+\frac{1}{2} \lambda \vcb^2 -
\lambda\vcb^2 \left[1-R_t e^{i\beta}\right] +
\ord(\lambda^7),
\ee
\be
V_{us}=\lambda+\ord(\lambda^7), \qquad
V_{cs}= 1-\frac{1}{2}\lambda^2-\frac{1}{8}\lambda^4 -\frac{1}{2} \vcb^2
 +\ord(\lambda^6),
\ee
\be
V_{tb}=1-\frac{1}{2} \vcb^2+\ord(\lambda^6),
\qquad
V_{td}=\lambda\vcb R_t e^{-i\beta}
+\ord (\lambda^7),
\ee
\begin{equation}
 V_{ts}= -\vcb +\frac{1}{2} \lambda^2 \vcb - 
\lambda^2 \vcb \left[1-R_t e^{i\beta}\right]
  +\ord(\lambda^6)~,
\end{equation}
where in order to simplify the notation we used $\lambda$ instead of $\vus$.

The CKM matrix determined in this manner and the corresponding unitarity 
triangle with 
\be\label{UUTR}
\bar\varrho=1-R_t\cos\beta, \qquad
\bar\eta=R_t\sin\beta
\ee
are universal in the class of MFV models. We will determine the universal
unitarity triangle (UUT) below.

\subsection{General Procedure in the SM}
After these general discussion let us concentrate first 
on the standard analysis of the UT within the SM. 
A very detailed description of this analysis with the participation of 
the leading experimentalists and theorists in this field 
can be found in  \cite{CERNCKM}. The relevant background can be found in 
\cite{Schladming,Erice}.
 
Setting $\lambda=\vus=0.224$, the analysis
proceeds in the following five steps:

{\bf Step 1:}

{}From  $b\to c$ transition in inclusive and exclusive 
leading $B$ meson decays
one finds $\vcb$ as given in (\ref{vcb}) and consequently the scale of the UT:
\begin{equation}\label{S1}
\vcb\quad \Longrightarrow\quad\lambda \vcb= \lambda^3 A~, 
\qquad (A=0.83\pm 0.02).
\end{equation}

{\bf Step 2:}

{}From  $b\to u$ transition in inclusive and exclusive $B$ meson decays
one finds $\vub$ as given in (\ref{v13}) and consequently using (\ref{2.94}) 
the side $CA=R_b$ of the UT:
\begin{equation}\label{rb}
\left| \frac{V_{ub}}{V_{cb}} \right|
 \quad\Longrightarrow \quad R_b=\sqrt{\bar\varrho^2+\bar\eta^2}=
4.35 \cdot \left| \frac{V_{ub}}{V_{cb}} \right|
\quad\Longrightarrow \quad
R_b=0.37\pm 0.04~.
\end{equation}

{\bf Step 3:}

{}From the experimental value of the CP-violating parameter $\varepsilon_K$ 
that describes the indirect CP violation in $K\to\pi\pi$ and 
the standard expression for box diagrams   
one derives  
the constraint on $(\bar \varrho, \bar\eta)$ \cite{WARN}
\begin{equation}\label{100}
\bar\eta \left[(1-\bar\varrho) A^2 \eta^{  QCD}_2 S_0(x_t)
+ P_c(\varepsilon) \right] A^2 \hat B_K = 0.187,
\end{equation}
where
$P_c(\varepsilon)=0.29\pm0.07$ \cite{charm} summarizes the contributions
of box diagrams with two charm quark exchanges and the mixed 
charm-top exchanges.  
The dominant term involving $\eta^{  QCD}_2=0.57\pm0.01$ 
\cite{BJW90} represents 
the box contributions with two top quark exchanges. $\hat B_K$ is 
a non-perturbative parameter for which the value is given below. 
As seen in fig.~\ref{L:10}, equation (\ref{100}) specifies 
a hyperbola in the $(\bar \varrho, \bar\eta)$
plane.

\begin{figure}[hbt]
  \vspace{0.10in} \centerline{
\begin{turn}{-90}
  \mbox{\epsfig{file=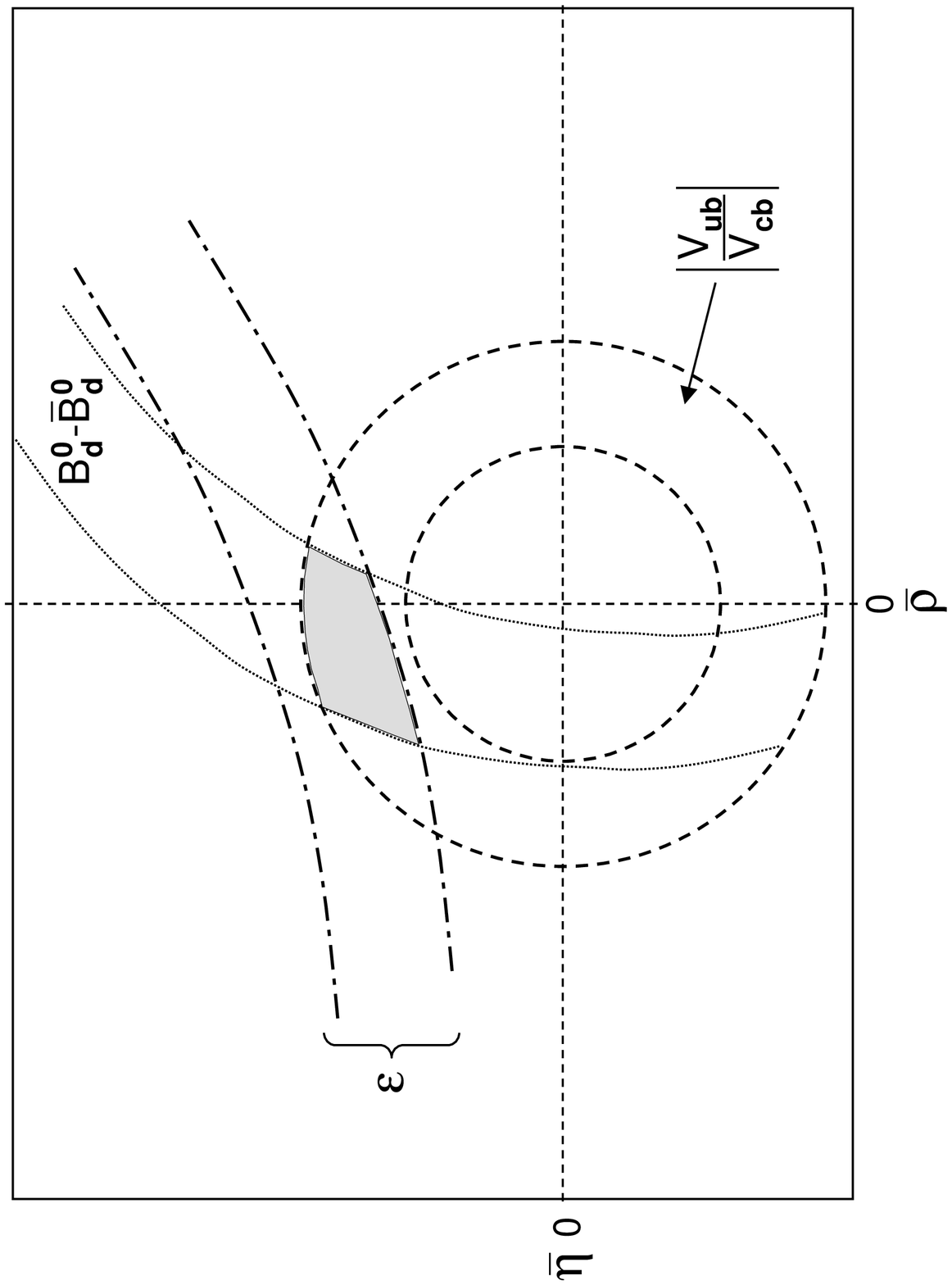,width=0.5\linewidth}}
\end{turn}
} \vspace{-0.18in}
\caption[]{Schematic determination of the Unitarity Triangle.
\label{L:10}}
 \end{figure}
{\bf Step 4:}

{}From the measured $\Delta M_d$ that represents the $B_d^0-\bar B_d^0$ 
mixing and the box diagrams with two top quark exchanges
the side $AB=R_t$ of the UT can be determined:
\begin{equation}\label{106}
 R_t= \frac{1}{\lambda}\frac{|V_{td}|}{\vcb} = 0.85 \cdot
\left[\frac{|V_{td}|}{7.8\cdot 10^{-3}} \right] 
\left[ \frac{0.041}{\vcb} \right],
\end{equation}
\begin{equation}\label{VT}
\vtd=
7.8\cdot 10^{-3}\left[ 
\frac{230\mev}{\sqrt{\hat B_{B_d}}F_{B_d}}\right]
\left[\frac{\Delta M_d}{0.50/{\rm ps}} \right ]^{0.5} 
\sqrt{\frac{0.55}{\eta_B^{  QCD}}}
\sqrt{\frac{2.40}{S_0(x_t)}}~.
\end{equation}
Here $\sqrt{\hat B_{B_d}}F_{B_d}$ is a non-perturbative parameter and 
$\eta_B^{  QCD}=0.55\pm0.01$ the QCD correction \cite{BJW90,UKJS}.
Moreover $\mtb(\mt)=(167\pm 5)$ GeV.
The constraint in the $(\bar\varrho,\bar\eta)$ plane coming from
this step is illustrated in fig.~\ref{L:10}.

{\bf Step 5:}

{}The measurement of  $\Delta M_s$
together with $\Delta M_d$  allows to determine $R_t$ in a different
manner:
\be\label{Rt}
R_t=0.90~\left[\frac{\xi}{1.24}\right] \sqrt{\frac{18.4/ps}{\Delta M_s}} 
\sqrt{\frac{\Delta M_d}{0.50/ps}},
\qquad
\xi = 
\frac{\sqrt{\hat B_{B_s}}F_{B_s} }{ \sqrt{\hat B_{B_d}}F_{B_d}}.
\ee
One should 
note that $\mt$ and $|V_{cb}|$ dependences have been eliminated this way
 and that $\xi$ should in principle 
contain much smaller theoretical
uncertainties than the hadronic matrix elements in $\Delta M_d$ and 
$\Delta M_s$ separately.  

The main uncertainties in these steps originate in the theoretical 
uncertainties in  $\hat B_K$ and 
$\sqrt{\hat B_d}F_{B_d}$ and to a lesser extent in $\xi$ 
\cite{CERNCKM}: 
\be\label{paramet}
\hat B_K=0.86\pm0.15, \quad  
\sqrt{\hat B_d}F_{B_d}=(235^{+33}_{-41})~{\rm MeV},
\quad \xi=1.24\pm 0.08~.
\ee
Also the uncertainties due to $\vub$ in step 2  
are substantial. The QCD sum rules results for  
the parameters in question are similar and can be found in 
\cite{CERNCKM}. 
Finally \cite{CERNCKM}
\be
\Delta M_d=(0.503\pm0.006)/{\rm ps}, \qquad 
\Delta M_s>14.4/{\rm ps}~~ {\rm at }~~ 95\%~{\rm C.L.}
\ee 
\subsection{The Angle {$\beta$} from {$B_d\to \psi K_S$}}
One of the highlights of the year 2002 were the considerably improved 
measurements of 
$\sin2\beta$ by means of the time-dependent CP asymmetry
\be\label{asy}
a_{\psi K_S}(t)\equiv -a_{\psi K_S}\sin(\Delta M_d t)=
-\sin 2 \beta \sin(\Delta M_d t)~.
\ee
The BaBar \cite{BaBar} and Belle \cite{Belle} collaborations find
\begin{displaymath}\label{sinb}
(\sin 2\beta)_{\psi K_S}=\left\{
\begin{array}{ll}
0.741\pm 0.067 \, \mbox{(stat)} \pm0.033 \, \mbox{(syst)} & \mbox{(BaBar)}\\
0.719\pm 0.074 \, \mbox{(stat)} \pm0.035  \, \mbox{(syst)} & \mbox{(Belle).}
\end{array}
\right.
\end{displaymath}
Combining these results with earlier measurements by CDF,
ALEPH and OPAL 
gives the grand average \cite{NIR02}
\be
(\sin 2\beta)_{\psi K_S}=0.734\pm 0.054~.
\label{ga}
\ee
This is a mile stone in the field of CP violation and in the tests of the
SM as we will see in a moment. Not only violation of this symmetry has been 
confidently established 
in the $B$ system, but also its size has been measured very accurately.
Moreover in contrast to the five constraints listed above, the determination 
of the angle $\beta$ in this manner is theoretically very clean.

\subsection{Unitarity Triangle 2003 (SM)}
We are now in a position to combine all these constraints in order to 
construct the unitarity triangle and determine various quantities of interest.
In this context the important issue is the error analysis of these formulae, 
in particular the treatment of theoretical uncertainties. In the 
literature the most popular are the 
Bayesian approach \cite{C00} and the frequentist approach \cite{FREQ}. 
For the PDG analysis see \cite{PDG}.
A critical comparison of these and other methods can be found in 
\cite{CERNCKM}.
I can recommend this reading. 

\begin{figure}[htb!]
\begin{center}
{\epsfig{figure=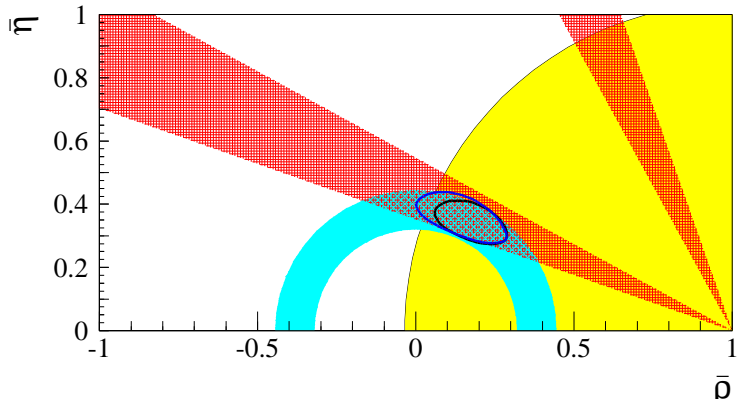,height=6cm}}
\caption[]{The allowed 95$\%$ regions in the 
$(\bar\varrho,\bar\eta)$ plane in the SM (narrower region) and in the 
MFV models (broader region) from the update of \cite{BUPAST}.
The individual 95$\%$ regions for the constraint from 
$\sin 2 \beta$, $\Delta M_s$ and $R_b$ are also shown.  
}
\label{fig:figmfv}
\end{center}
\end{figure}

In fig.~\ref{fig:figmfv} we show the result of the recent update of 
an analysis in collaboration 
with Parodi and Stocchi \cite{BUPAST} that uses 
the Bayesian approach. 
The allowed region for $(\bar\varrho,\bar\eta)$ 
is the area inside the smaller ellipse.
We observe that the region
$\bar\varrho<0$ is disfavoured by the lower bound on
$\Delta M_s$.
It is clear
from this figure that the measurement of $\Delta M_s$
giving $R_t$ through (\ref{Rt}) will have a large impact
on the plot in fig.~\ref{fig:figmfv}.  

The ranges for various quantities that result from this analysis 
are given in the SM column of table~\ref{mfv}. The UUT column 
will be discussed soon. The SM results follow from 
the five steps listed above and (\ref{ga}) implying an impressive 
precision on the angle $\beta$:
\begin{equation}\label{TOT}
(\sin 2\beta)_{\rm tot}=0.705^{+0.042}_{-0.032}, \qquad 
 \beta=(22.4 \pm 1.5)^\circ~.
\end{equation}
On the other hand $(\sin 2\beta)_{ind}$ obtained by means of 
the five steps only is found to be 
\cite{BUPAST} 
\be
(\sin 2\beta)_{ind}=0.685 \pm 0.052,
\label{ind}
\ee
demonstrating an excellent agreement (see also fig.~\ref{fig:figmfv}) 
between the direct measurement in (\ref{ga})
and the standard analysis of
the UT within the SM.
This gives a strong indication that the CKM matrix is very likely 
the dominant source of CP violation in flavour violating decays and gives 
a support to the MFV idea.
In order to be sure whether this is indeed the case, other theoretically
clean quantities have to be measured. In particular the angle $\gamma$ 
that is more sensitive to new physics contributions than $\beta$.
We refer to \cite{Schladming} and \cite{REV} for  reviews of the methods 
relevant for this determination.

\begin{table}[htb!]
\begin{center}
\caption[]{ \small { Values for different quantities from 
the update of \cite{BUPAST}.
$\lambda_t=V_{ts}^*V_{td}$.}}
\begin{tabular}{|c|c|c|}
\hline
  Strategy       &               UUT              &            SM         \\
  $\bar {\eta}$  &       ~ ~ 0.361 $\pm$ 0.032~~  &   ~~0.341 $\pm$ 0.028~~  \\ 
%                 &                                &                        \\
  $\bar {\varrho}$  &       ~~0.149 $\pm$ 0.056~~      &   ~~ 0.178 $\pm$ 0.046~~   \\
%                 &                                &                         \\
  $\sin 2\beta$  &  ~~0.715 $^{+0.037}_{-0.034}$~~   & ~~0.705 $^{+0.042}_{-0.032}$~~ \\
%                 &                                &                         \\
  $\sin 2\alpha$ &     ~~0.03 $\pm$ 0.31~~         & ~~ --0.19 $\pm$ 0.25~~           \\
%                 &                                &                       \\
  $\gamma$       &   ~~$ (67.5 \pm 8.9)^\circ$~~    &   ~~$ (61.5 \pm 7.0)^\circ $~~\\
%                 &                                &                       \\
  $R_b$          &       ~~0.393 $\pm$ 0.025~~      & ~~0.390 $\pm$ 0.024~~   \\
%                 &                                &                        \\
  $R_t$          &       ~~$0.925\pm 0.060$~~        &    ~~$0.890 \pm 0.048 $~~  \\
%                 &                                &                        \\
  $\Delta M_s$ ($ps^{-1}$)   & ~~17.3$^{+2.1}_{-1.3}$~~   &  ~~18.3$^{+1.7}_{-1.5}$~~       \\
%                 &                                &                       \\
$\vtd~(10^{-3})$ &        ~~8.61 $\pm$ 0.55~~       &      ~~8.24 $\pm$ 0.41~~       \\
%                 &                                &                         \\
~~${\rm Im} \lambda_t$ ($10^{-4}$)~~  & ~~1.39 $\pm$ 0.12~~    &    ~~1.31 $\pm$ 0.10~~    \\
%                 &                                &               \\
\hline
\end{tabular}
\label{mfv}
\end{center}
\end{table}

\subsection{Unitarity Triangle 2003 (MFV)}
In a general MFV model the formulae (\ref{S1})--(\ref{ga}) still apply with
$S_0(x_t)$ replaced by the master function $S(v)$. In particular as
emphasized in \cite{BF01}, $S(v)$ could be negative resulting in
$\bar\eta<0$. As found in \cite{BF01,AMGIISST} this case is disfavoured but 
not yet excluded. Here we will only consider $S(v)>0$. We recall that in the 
absence of new CP violating phases, (\ref{ga}) determines a universal angle 
$\beta$ in the MFV models. Moreover we note that not only (\ref{ga}) but 
also (\ref{S1}), (\ref{rb}) and (\ref{Rt}) do not involve $S(v)$ and are 
universal within the MFV models. Using only them allows to construct a universal 
unitarity triangle (UUT) common to all these models \cite{UUT}. The apex 
of the UUT is positioned within the larger ellipse 
in figure \ref{fig:figmfv}  
as obtained recently in an update of 
\cite{BUPAST}. The results for various quantities of 
interest related to this UUT are collected in table \ref{mfv}.
A similar analysis has been done in \cite{AMGIISST}.

It should be stressed that any MFV model that is inconsistent with the 
broader allowed region in figure \ref{fig:figmfv} and 
the UUT column in table \ref{mfv} is ruled out. 
We observe that there is little room for MFV models that in their predictions 
for UT differ significantly from the SM. It is also clear that to 
distinguish the SM from the MFV models on the 
basis of the analysis of the UT presented above, will require 
considerable reduction of 
theoretical uncertainties.
Therefore for the near future the most precise determination of the UUT will 
come from $\sin 2\beta$ measured through 
$a_{\psi K_S}$ and the ratio $\Delta M_s/\Delta M_d$ as advocated in Section 
3.2. 
\section{Relations from Minimal Flavour Violation}
\subsection{Preliminaries}
We have seen that an UUT could be constructed.
In this construction the relation between $R_t$ and the ratio 
$\Delta M_d/\Delta M_s$, that does not depend on $F_r$, played an important 
role. 
It is clear from the tables in Sections 2.5 and 2.8 that
there are other interesting  relations between branching ratios and 
various observables 
which similarly to UUT are independent of $F_r$. These relations are 
very important for testing the general concept of MFV. 
In order to illuminate the origin of the MFV relations in question we will 
now list the formulae for most important decays in the MFV models. 
Because of space limitations our presentation will be reduced to the minimum.
Details on these formulae 
can be found in \cite{Erice,AJBLH,BSW02,BPSW} and references therein.
The numerical constants in the formulae below correspond to \cite{PDG} 
and 
\be
\sin^2\theta_w=0.231, \qquad \alpha=\frac{1}{128}, \qquad \lambda=0.224
\ee
with the first two given in the $\overline{MS}$ scheme. They differ slightly 
from those in \cite{Erice,AJBLH,BSW02,BPSW}.
\subsection{Basic Formulae}

\subsubsection{The \boldmath{$\varepsilon_K$} Constraint}
Similarly to (\ref{100}) we have
\begin{equation}\label{100M}
\bar\eta \left[(1-\bar\varrho) A^2 \eta^{  QCD}_2 S(v)
+ P_c(\varepsilon) \right] A^2 \hat B_K = 0.187,
\end{equation}
where
$P_c(\varepsilon)=0.29\pm0.07$ and $\eta^{QCD}_2=0.57\pm0.01$ as before.

\subsubsection{\boldmath{$B^0_{d,s}-\bar B^0_{d,s}$} Mixing}
 Within the MFV models $\Delta M_{s,d}$ 
are given as follows ($q=d,s$) \cite{BBL}
\begin{equation}\label{DMQMFV}
\Delta M_q = \frac{G_{\rm F}^2}{6 \pi^2} \eta_B^{  QCD} m_{B_q} 
(\hat B_{q} F_{B_q}^2 ) \mw^2  
|V^\ast_{tb}V_{tq}|^2 S(v),
\end{equation}
where
$\eta_B^{  QCD}=0.55\pm0.01$ \cite{BJW90,UKJS},
impling (\ref{106})--(\ref{Rt}) with $S_0(x_t)$ 
in (\ref{VT}) replaced by $S(v)$.

\subsubsection{\boldmath{$K\to\pi\nu\bar\nu$}}
The rare decays $\kpn$ and $\klpn$  
proceed through $Z^0$-penguin
 and box diagrams. As the required hadronic matrix elements can be extracted 
from the leading semileptonic decays and other long distance contributions 
turn out to be negligible \cite{Gino03}, 
the relevant branching ratios can be 
computed
to an exceptionally high degree of precision \cite{BB,BB98,MU98}. 

The basic formulae for the branching ratios are given in MFV as follows
\begin{equation}\label{bkpn}
Br(\kpn)=4.78\cdot 10^{-11}\cdot\left[\left(\IM F_t \right)^2+
\left(\RE F_c +\RE F_t\right)^2 \right]~,
\end{equation}
\begin{equation}\label{bklpn}
Br(K_{\rm L}\to\pi^0\nu\bar\nu)=2.09\cdot 10^{-10}\cdot
\left(\IM F_t  \right)^2~,
\end{equation}
where
\be 
F_c={\lambda_c\over\lambda}P_c(X), \qquad 
F_t={\lambda_t\over\lambda^5}X(v), \qquad \lambda_i=V^\ast_{is}V_{id}
\ee
and 
$P_c(X)=0.39\pm 0.06$ results from the internal charm 
contribution \cite{BB,BB98}.

Imposing all existing constraints on the CKM matrix one finds 
\cite{Gino03,KENDAL}
\begin{equation}\label{kpnr}
Br(\kpn)=
(7.7 \pm 1.1)\cdot 10^{-11},\qquad Br(\klpn)=
(2.6 \pm 0.5)\cdot 10^{-11}  
\end{equation}
where the errors come dominantly from the uncertainties in the CKM
parameters. 
 This should be compared respectively with the results of  
AGS E787 collaboration \cite{Adler97} and KTeV \cite{KTeV00X}
\be\label{kp01}
Br(K^+ \rightarrow \pi^+ \nu \bar{\nu})=
(15.7^{+17.5}_{-8.2})\cdot 10^{-11}, \qquad
Br(\klpn)<5.9 \cdot 10^{-7}\,.
\end{equation}

\subsubsection{\boldmath{$B\to X_{s,d}\nu\bar\nu$}}
Theoretically clean \cite{BB,BI98} 
are also the inclusive decays 
$B\to X_{s,d}\nu\bar\nu$ for which the branching ratios read
\begin{equation}\label{bbxnn}
Br(B\to X_q\nu\bar\nu)= 1.58\cdot 10^{-5}
\left[\frac{Br(B\to X_c e\bar\nu)}{0.104}\right]
\left[\frac{0.54}{f(z)}\right]
\frac{|V_{tq}|^2}{|V_{cb}|^2}X^2(v)
\end{equation}
where $f(z)$ is a phase space factor for $B\to X_c e\bar\nu$.
The SM expectation 
\be
Br(B\to X_s\nu\bar\nu)=(3.5\pm 0.5)\cdot 10^{-5}
\ee
is to be compared with the $90\%$ C.L ALEPH upper bound
$6.4\cdot 10^{-4}$. The exclusive channels are less clean but 
experimentally easier accessible with the $90\%$ C.L BaBar upper bound
of $7.0\cdot 10^{-5}$. 
\subsubsection{\boldmath{$K_L\to \mu^+\mu^-$}}
The short distance contribution to the dispersive part of $\kmm$ is given 
by \cite{BB,BB98} 
\begin{equation}\label{bklm}
Br(\kmm)_{\rm SD}=1.95\cdot 10^{-9}\left[\frac{\relc}{\lambda}P_c(Y)+
\frac{\relt}{\lambda^5} Y(v)\right]^2,
\end{equation}
where $P_c(Y) = 0.121\pm 0.012$ \cite{BB98}.  
Unfortunately due to long distance contributions to the dispersive part of 
$K_L\to \mu^+\mu^-$, the extraction of $Br(\kmm)_{\rm SD}$ from the data is 
subject to considerable uncertainties \cite{ISIDORI,EDUARDO}. 
While the chapter on this extraction is certainly not closed, let us quote 
the estimate of \cite{ISIDORI} that reads 
 \begin{equation}\label{SD}
Br(\kmm)_{\rm SD}\le 2.5\cdot 10^{-9}~
\end{equation}
to be compared with $Br(\kmm)_{\rm SD}=(0.8\pm 0.3) \cdot 10^{-9}$ in the SM.

On the other hand the parity violating muon polarization asymmetry 
$\Delta_{\rm LR}$ in $K^+\to\pi^+\mu^+\mu^-$ is
substantially cleaner \cite{Savage}. It is given at 
NLO level by \cite{BB94}
\be
|\Delta_{\rm LR}|=  r \cdot 1.78\cdot 10^{-3}
\left|\frac{\relc}{\lambda}P_c(Y)+
\frac{\relt}{\lambda^5} Y(v) \right|,
\ee
where $r$ is a phase factor that may depend on various experimental cuts.

\subsubsection{\boldmath{$B_q\to \mu\bar\mu$}}
Next, the branching ratios for the rare decays $B_q\to\mu^+\mu^-$ are
given by
\begin{equation}\label{bbll}
Br(B_q\to \mu^+\mu^-)=\tau(B_q)\frac{G^2_{\rm F}}{\pi}\eta_Y^2
\left(\frac{\alpha}{4\pi\sin^2\theta_{W}}\right)^2 F^2_{B_q}m^2_\mu m_{B_q}
|V^\ast_{tb}V_{tq}|^2 Y^2(v),
\end{equation}
where
$\eta_Y=1.012$ \cite{BB98} are the short distance 
QCD corrections evaluated using $m_t\equiv\overline{m}_t(m_t)$. 
In writing (\ref{bbll}) we have neglected the
terms $\ord(m_\mu^2/m_{B_q}^2)$ in the phase space factor. 
In the SM one finds \cite{AJB03} (see Section 4.4)
\be\label{Results}
 Br(B_{s}\to\mu\bar\mu)=(3.42\pm 0.54)\cdot 10^{-9}, \qquad
Br(B_{d}\to\mu\bar\mu)=(1.00\pm 0.14)\cdot 10^{-10}
\ee
where to reduce the hadronic uncertainties the experimental data for 
$\Delta M_{d}$ and as an example $\Delta M_s=(18.0\pm0.5)/{\rm ps}$ 
have been used. This should be compared respectively with the $90\%$ C.L.
 bounds from 
CDF(D0) and Belle \cite{Nakao,BelleBd}
\be
Br(B_{s}\to\mu\bar\mu)< 9.5~(16) \cdot 10^{-7}, \qquad
Br(B_{d}\to\mu\bar\mu)< 1.6 \cdot 10^{-7}.
\ee

\subsubsection{ \boldmath{$K_L\to\pi^0 e^+ e^-$}} 
The rare decay $K_L\to\pi^0 e^+e^-$ is dominated by CP-violating 
contributions. 
It has been recently reconsidered within the SM \cite{BI03} in view 
of the most recent NA48 data  on $K_S\to\pi^0 e^+e^-$
and $K_L\to \pi^0\gamma\gamma$ \cite{NA48KL} that allow a much better 
evaluation of the indirectly (mixing) CP-violating and CP-conserving  
contributions. 
The directly CP-violating contribution has been know already at NLO for some 
time \cite{BLMM}. 
The CP-conserving part is found to be below $3\cdot 10^{-12}$ \cite{BI03}.

Generalizing the formula (33) in \cite{BI03} to MFV models we obtain 
\begin{equation}\label{9a}
Br(K_{\rm L} \to \pi^0 e^+ e^-)_{\rm CPV} =
10^{-12}\cdot\left[C_{\rm mix}+
C_{\rm int}\left(\frac{\IM\lambda_t}{10^{-4}}\right) +
C_{\rm dir}\left(\frac{\IM\lambda_t}{10^{-4}}\right)^2\right],
\ee
where
\be
C_{\rm mix}=(15.7\pm 0.3)|a_s|^2, \qquad |a_s|=1.08^{+0.26}_{-0.21},
\ee
\be
C_{\rm dir}=6.2\cdot 10^{-2}
(\tilde y_{7A}^2 + \tilde y_{7V}^2)\,,\qquad
C_{\rm int}=0.34\, \tilde y_{7V}\,\sqrt{C_{\rm mix}}.
\end{equation}
Here
\begin{equation}\label{y7vpbe}
\tilde{y}_{7V} =
P_0 + \frac{Y(v)}{\sin^2\theta_{ w}} - 4 Z(v)+ P_E E(v)~,
\qquad
\tilde{y}_{7A}=-\frac{1}{\sin^2\theta_{w}} Y(v),
\end{equation}
where
$\IM \lambda_t = \IM (V_{td} V^*_{ts})$, 
 $P_0=2.89\pm0.06$  \cite{BLMM} and 
$P_E$ is $\ord(10^{-2})$. Consequently the last
term in $\tilde{y}_{7V}$ can be neglected. 
The effect of new physics 
contributions is mainly 
felt in $\tilde{y}_{7A}$ as the corresponding contributions in 
$\tilde{y}_{7V}$ cancel each other to a large extent.

The present experimental bound from KTeV \cite{KTEVKL}
\be 
Br(K_{\rm L} \to \pi^0 e^+ e^-)< 2.8\cdot 10^{-10}\qquad (90\% {\rm C.L.})
\ee
should be compared with the SM prediction \cite{BI03}
\be
Br(K_{\rm L} \to \pi^0 e^+ e^-)_{\rm SM}= (3.2^{+1.2}_{-0.8})\cdot 10^{-11}~.
\ee

\subsubsection{\boldmath{$\epe$}}
The
formula for the CP-violating ratio $\epe$ of \cite{BJ03} generalizes to 
the arbitrary MFV
 model
as follows: 
\be \frac{\varepsilon'}{\varepsilon}= \IM\lambda_t
\cdot F_{\varepsilon'}(v)
\label{epeth}
\ee
where
\be
F_{\varepsilon'}(v) =P_0 + P_X \, X(v) + 
P_Y \, Y(v) + P_Z \, Z(v)+ P_E \, E(v)~.
\label{FE}
\ee
The numerical values of the coefficients $P_i$ can be found in 
\cite{BJ03}. They depend strongly on the hadronic matrix elements of the 
relevant operators and the value of $\alpha_s$. 
For instance for the non-perturbative parameters $R_6=1.2$ and $R_8=1.0$ in 
\cite{BJ03} and $\alpha_s(M_Z)=0.119$ one has
\be\label{PI1}
P_0=19.5, \qquad P_X=0.6, \qquad P_Y=0.5, \qquad P_Z=-12.4, \qquad P_E=-1.6
\ee
with $P_0$ and $P_Z$ originating dominantly in the matrix elements of the 
QCD penguin $Q_6$ and the electroweak penguin $Q_8$, respectively.
On the other for $R_6=1.6$, $R_8=0.8$  and $\alpha_s(M_Z)=0.121$ one has
\be\label{PI2}
P_0=28.6, \qquad P_X=0.6, \qquad P_Y=0.6, \qquad P_Z=-10.3, \qquad P_E=-2.7~.
\ee

On the experimental side the world
average based on the latest results from NA48 \cite{NA48} and KTeV
\cite{KTeV}, and previous results from NA31  and E731,
 reads
\begin{equation}
  \label{eps}
  \epe=(16.6\pm 1.6) \cdot 10^{-4} \qquad\qquad (2003)~.
\end{equation}
 While several analyses of recent years
within the SM find results that are compatible with 
(\ref{eps}) it is fair to say, in view of large hadronic uncertainties 
in the coefficients $P_i$,
that the chapter on the theoretical calculations of
$\epe$ is far from being closed. 
For instance with $m_t=167\gev$ and $\IM\lambda_t=1.44\cdot 10^{-4}$ (see an 
example in Section 5.5) one finds in the SM $\epe=17.4 \cdot 10^{-4}$ and 
$32.2 \cdot 10^{-4}$ for (\ref{PI1}) and (\ref{PI2}), respectively.
The most recent analysis with 
the relevant references is given in \cite{BJ03}.

\subsubsection{\boldmath{$B\to X_s\gamma$} and 
\boldmath{$B\to X_s~{\rm gluon}$} }
These decays are governed by
the {\it magnetic--penguin} operators
\begin{equation}\label{O6B}
Q_{7\gamma}  =  \frac{e}{8\pi^2} m_b \bar{s}_\alpha \sigma^{\mu\nu}
          (1+\gamma_5) b_\alpha F_{\mu\nu},\qquad            
Q_{8G}     =  \frac{g_s}{8\pi^2} m_b \bar{s}_\alpha \sigma^{\mu\nu}
   (1+\gamma_5)T^a_{\alpha\beta} b_\beta G^a_{\mu\nu}  
\end{equation}
originating in the diagrams of fig.~\ref{fig:fdia}d with an on-shell
photon and gluon, respectively.
Their Wilson coefficients are strongly affected by QCD corrections 
\cite{Bert,Desh} coming 
dominantly from the mixing of $Q_{7\gamma}$ and $Q_{8G}$
with
current--current operators.

In the leading logarithmic approximation one has 
\begin{equation}\label{main}
{Br(B \to X_s \gamma)}=2.88\cdot 10^{-3}
\left[\frac{Br(B\to X_c e\bar\nu)}{0.104}\right]
\left[\frac{0.54}{f(z)}\right]
\frac{|V_{ts}|^2}{|V_{cb}|^2}
|C^{(0){\rm eff}}_{7}(\mu_b)|^2\,,
\end{equation}
where $f(z)$ 
is the phase space factor in $Br(B \to X_c e \bar{\nu}_e)$.
The Wilson coefficient $C^{(0){\rm eff}}_{7}(\mu_b)$
is given by
\be
C^{(0){\rm eff}}_{7\gamma}(\mu_b) = -0.348\, D'(v)
-0.042\, E'(v) -0.158
\label{eq:C7geffnum}
\ee
where we have set $\mu_b = 5\gev$ and 
$\as^{(5)}(\mz)=0.118$. The last term in this formula comes from the mixing 
with the current--current operator $Q_2$ and the coefficients in front 
of the master functions come from the renormalization group analysis. The 
small coefficient in front of $E'(v)$ makes this function subleading in
this decay.

The corresponding NLO formulae that include also higher order 
electroweak effects \cite{GAHA} are very complicated and can be found in 
\cite{Gambino:2001ew}.
As reviewed in \cite{AJBMIS,ALIMIS,HURTH}, many groups contributed to 
obtain these NLO results, in particular Christoph Greub, Mikolaj Misiak 
and their collaborators. See also \cite{Gambino:2003zm}.

On the 
experimental side the world average resulting from the data by CLEO, ALEPH, 
BaBar and Belle reads \cite{CERNCKM}
\be\label{bsgexp}
Br(B\to X_s\gamma)_{E_\gamma> 1.6{\rm GeV}}=
 (3.28^{+0.41}_{-0.36})\cdot 10^{-4}~.
\ee
It agrees well with the SM result \cite{Gambino:2001ew}
\be\label{bsgth}
Br(B\to X_s\gamma)^{\rm {SM}}_{E_\gamma> 1.6{\rm GeV}}
= (3.57\pm 0.30)\cdot 10^{-4}~.
\ee

The $B\to X_s~{\rm gluon}$ decay is dominated by $Q_{8G}$ with
$C^{(0){\rm eff}}_{8G}$ given by
\be
C^{(0){\rm eff}}_{8G}(\mu_b) =
-0.364\, E'(v) - 0.074 
\label{eq:C8Geffnum}
\ee
with the last term representing QCD renormalization group effect.
The NLO corrections have been calculated in \cite{GL00}. Unfortunately, 
the remaining 
strong renormalization scale dependence in the resulting branching ratio and 
the difficulty in extracting it from the experiment, make these results 
not yet useful at present.
   
\subsubsection{\boldmath{$B\to X_s\mu^+\mu^-$} and  \bf{$A_{FB}(\hat s)$}}
This decay is dominated by the operators
\begin{equation}\label{Q9V}
Q_{9V}    = (\bar{s} b)_{V-A}  (\bar{\mu}\mu)_V\,,         
\qquad
Q_{10A}  =  (\bar{s} b)_{V-A}  (\bar{\mu}\mu)_A\,.
\end{equation}
 They are generated through the electroweak
penguin diagrams of fig.~\ref{fig:fdia}f 
 and the related box diagrams are needed mainly
to keep gauge invariance. At low
\begin{equation} \label{invleptmass}
\hat s = \frac{(p_{\mu^+} + p_{\mu^-})^2}{\mb^2},
\end{equation}
also the magnetic operator $Q_{7\gamma}$ plays a significant role.

At the NLO level \cite{Mis:94,BuMu:94} the invariant dilepton mass spectrum
is given by
\be \label{rateee}
\frac{{d}/{d\hat s} \, 
\Gamma (b \to s \mu^+\mu^-)}{\Gamma
(b \to c e\bar\nu)} = \frac{\alpha^2}{4\pi^2}
\left|\frac{V_{ts}}{V_{cb}}\right|^2 \frac{(1-\hat s)^2}{f(z)\kappa(z)}
U(\hat s)
\ee
where
\be\label{US} 
U(\hat s)=
(1+2\hat s)\left(|\Ctilde_9^{\rm eff}(\hat s)|^2 + |\Ctilde_{10}|^2\right) + 
4 \left( 1 + \frac{2}{\hat s}\right) |C_{7\gamma}^{(0){\rm eff}}|^2 + 12
C_{7\gamma}^{(0){\rm eff}} \ \RE\,\Ctilde_9^{\rm eff}(\hat s) 
\ee
and $\Ctilde_9^{\rm eff}(\hat s)$ is a function of $\hat s$ that depends on
the Wilson coefficient $\Ctilde_9$ and includes also contributions 
from four quark operators. Explicit formula can be found in
\cite{Mis:94,BuMu:94}.

The Wilson coefficients $\Ctilde_9$ and  $\Ctilde_{10}$ are given as follows
\begin{equation}\label{C9tilde}
\Ctilde_9(\mu)  =  
P_0 + \frac{Y(v)}{\sin^2\theta_{w}} -4 Z(v) +
P_E E(v), \qquad
\tilde C_{10}(\mu) = - \frac{Y(v)}{\sin^2\theta_{w}}
\end{equation}
with
$P_0 = 2.60\pm 0.25$ in the NDR scheme and $P_E=\ord(10^{-2})$.
Note the great similarity to (\ref{y7vpbe}). 
 $\tilde C_{9}$ and
$\tilde C_{10}$ are defined by 
\begin{equation} \label{C10}
C_{9V}(\mu) = \frac{\alpha}{2\pi} \tilde C_9(\mu), \qquad
C_{10A}(\mu) =  \frac{\alpha}{2\pi} \tilde C_{10}(\mu).
\ee

Of particular interest is the Forward-Backward asymmetry in 
$B\to X_s\mu^+\mu^-$. It becomes non-zero only at the NLO level and 
is given in this approximation by 
\cite{AMM}
\be\label{ABF}
A_{\rm FB}(\hat s)= -3 \tilde C_{10}
\frac{\left[\hat s \RE\,\Ctilde_9^{\rm eff}(\hat s)
+2 C_{7\gamma}^{(0){\rm eff}}\right]}
{U(\hat s)}
\ee
with $U(\hat s)$ given in (\ref{US}).
Similar 
to the case of exclusive 
decays \cite{Burdman}, the asymmetry $A_{\rm FB}(\hat s)$ vanishes at 
$\hat s=\hat s_0$ that in the case of the inclusive decay considered 
is determined through 
\be\label{ZERO}
\hat s_0 \RE\,\Ctilde_9^{\rm eff}(\hat s_0)+2 C_{7\gamma}^{(0){\rm eff}}=0.
\ee
$A_{\rm FB}(\hat s)$ and the value of $\hat s_0$ 
are sensitive to short distance physics and 
subject to only very small non-perturbative uncertainties. Consequently, they 
are 
particularly useful quantities to test the physics beyond the SM. 

The calculations of $A_{\rm FB}(\hat s)$ and of $\hat s_0$ in the SM have
recently been 
done including NNLO corrections \cite{NNLO1,NNLO2} that turn out to 
be significant. 
In particular they shift the NLO value of $\hat s_0$ from $0.142$ to
$0.162$ at NNLO.

The most recent reviews summarizing the theoretical status can be found 
in \cite{HURTH,Ali:2002jg}. 
On the experimental side the  Belle and BaBar collaborations \cite{Kaneko:2002mr}
 reported the observation of 
this decay and of the $X_se^+e^-$ channel. 
The $90\%$ C.L. ranges extracted from these papers \cite{HIKR} 
read
\be
3.5\cdot 10^{-6}\le Br(B\to X_s\mu^+\mu^-)\le 10.4\cdot 10^{-6}~, 
\ee
\be
2.8\cdot 10^{-6}\le Br(B\to X_s\mu^+\mu^-)\le 8.8\cdot 10^{-6}.
\ee

\subsection{Model Independent Relations}
We will now list a number of relations between various observables that 
do not depend on the functions $F_r(v)$ and consequently are universal 
within the class of MFV models.

{\bf 1.} From (\ref{DMQMFV}), (\ref{bbxnn}) and (\ref{bbll}) 
we find \cite{BBSIN} 
\be\label{dmsdmd}
\frac{\Delta M_d}{\Delta M_s}=
\frac{m_{B_d}}{m_{B_s}}
\frac{\hat B_{d}}{\hat B_{s}}\frac{F^2_{B_d}}{F^2_{B_s}}
\left|\frac{V_{td}}{V_{ts}}\right|^2
\end{equation}
\begin{equation}\label{bxnn}
\frac{Br(B\to X_d\nu\bar\nu)}{Br(B\to X_s\nu\bar\nu)}=
\left|\frac{V_{td}}{V_{ts}}\right|^2
\end{equation}
\begin{equation}\label{bmumu}
\frac{Br(B_d\to\mu^+\mu^-)}{Br(B_s\to\mu^+\mu^-)}=
\frac{\tau({B_d})}{\tau({B_s})}\frac{m_{B_d}}{m_{B_s}}
\frac{F^2_{B_d}}{F^2_{B_s}}
\left|\frac{V_{td}}{V_{ts}}\right|^2
\end{equation}
that all can be used to determine $|V_{td}/V_{ts}|$ without the knowledge
of $F_r(v)$ \cite{UUT}. In particular, as already emphasized in Section
3,  the relation (\ref{dmsdmd}) 
will offer
after the measurement of $\Delta M_s$ a  powerful determination of the 
length of one side of the unitarity triangle, denoted usually by $R_t$.

Out of these three ratios the cleanest 
is (\ref{bxnn}), which is essentially free
of hadronic uncertainties \cite{BI98}. Next comes (\ref{bmumu}), involving
$SU(3)$ breaking effects in the ratio of $B$-meson decay constants.
Finally, $SU(3)$ breaking in the ratio 
$\hat B_{B_d}/\hat B_{B_s}$ enters in addition in (\ref{dmsdmd}). These 
$SU(3)$ breaking effects should eventually be calculable with
high precision from lattice QCD.

Eliminating $|V_{td}/V_{ts}|$ from the three relations above allows 
to obtain three relations between observables that are universal within the
MFV models. In particular 
from (\ref{dmsdmd}) and (\ref{bmumu}) one finds \cite{AJB03} 
\be\label{R1}
\frac{Br(B_{s}\to\mu\bar\mu)}{Br(B_{d}\to\mu\bar\mu)}
=\frac{\hat B_{d}}{\hat B_{s}}
\frac{\tau( B_{s})}{\tau( B_{d})} 
\frac{\Delta M_{s}}{\Delta M_{d}},
\ee
that does not 
involve $F_{B_q}$ and consequently contains 
substantially smaller hadronic uncertainties than the formulae considered 
above. It involves
only measurable quantities except for the ratio $\hat B_{s}/\hat B_{d}$
that is known already now from lattice calculations 
with respectable precision \cite{CERNCKM}:
\be\label{BBB}
\frac{\hat B_{s}}{\hat B_{d}}=1.00\pm 0.03, \qquad
\hat B_{d}=1.34\pm0.12, \qquad \hat B_{s}=1.34\pm0.12~.
\ee
With the future precise measurement of $\Delta M_s$, the formula (\ref{R1}) 
will give a very precise prediction for the ratio 
of the branching ratios $Br(B_{q}\to\mu\bar\mu)$.

{\bf 2.} Next, combining (\ref{bkpn}) and (\ref{bklpn}), it is possible to derive a very 
accurate formula for $\sin 2\beta$ that depends only on the 
$K\to\pi\nu\bar\nu$ branching 
ratios and $P_c(X)$ \cite{BBSIN}:
\be\label{sin2bnunu}
\sin2\beta= \frac{2 r_s}{1+r_s^2}, \qquad 
r_s=\sqrt{\sigma}{\sqrt{\sigma(B_1-B_2)}-P_c(X)\over\sqrt{B_2}}\,,
\ee 
where $\sigma=1/(1-\lambda^2/2)^2$ 
and we have assumed $X>0$. The corresponding formula valid also for $X<0$ is 
given in \cite{BF01}.
Here we have defined the ``reduced'' branching ratios
\begin{equation}\label{b1b2}
B_1={Br(\kpn)\over 4.78\cdot 10^{-11}},\qquad
B_2={Br(\klpn)\over 2.09\cdot 10^{-10}}.
\end{equation}
It should be stressed that $\sin 2\beta$ determined this way depends
only on two measurable branching ratios and on 
$P_c(X)$ which is completely calculable in perturbation theory.
Consequently this determination is free from any hadronic
uncertainties and its accuracy can be estimated with a high degree
of confidence. 
With measurements of $Br(\kpn)$ and $Br(\klpn)$
with $10\%$ accuracy a determination of $\sin 2\beta$ with an error of 
$0.05$ is possible.

Moreover, as in MFV models there are no phases beyond the CKM phase, 
 we also expect
\be\label{R7}
(\sin 2\beta)_{\pi\nu\bar\nu}=(\sin 2\beta)_{J/\psi K_S}, 
\qquad
(\sin 2\beta)_{\phi K_S}\approx (\sin 2\beta)_{J/\psi K_S}.
\ee
with the accuracy of the last relation at the level of a few percent 
\cite{Worah}.
The confirmation of these two relations would be a very important test for the 
MFV idea. Indeed, in $K\to\pi\nu\bar\nu$ the phase $\beta$ originates in 
the $Z^0$ penguin diagram, whereas in the case of $a_{J/\psi K_S}$ in 
the $B^0_d-\bar B^0_d$ box diagram. In the case of the asymmetry 
$a_{J/\phi K_S}$ it originates also in $B^0_d-\bar B^0_d$ box diagrams 
but the second relation in (\ref{R7}) could be spoiled by new physics 
contributions in the decay amplitude for $B\to \phi K_S$ that is
non-vanishing only at the one loop level.

An important consequence of (\ref{sin2bnunu}) and (\ref{R7}) is the following
one. 
For a given $\sin 2\beta$ extracted from $a_{J/\psi K_S}$ and $Br(\kpn)$ 
only two values of 
$Br(\klpn)$, corresponding to two signs of $X$, are possible 
in the full class of MFV models, independent of any new parameters 
present in these models \cite{BF01}. 
Consequently, measuring $Br(\klpn)$ will 
either select one of these two possible values or rule out all MFV models.
The present experimental bound on $Br(\kpn)$ and $\sin 2\beta\le 0.80$
imply in this manner an absolute upper bound 
$Br(\klpn)<4.9 \cdot 10^{-10}~(90\%~{\rm C.L.})$ 
\cite{BF01} in the MFV models that is by a factor of three stronger than 
the model independent bound \cite{GRNIR}  from isospin symmetry.
However, as we will see in Section 5.5, even stronger bound on
$Br(\klpn)$ can be obtained once the data on $B\to X_s\mu^+\mu^-$ are taken 
into account.

\begin{figure}[hbt]

  \centering 
 \psfragscanon
  \psfrag{bsgammabsgammabsgamma}{ \shortstack{\\ \\ $(Br(B\to
 X_s\gamma)\times 10^4)^\frac12$ ${}$}}
  \psfrag{hats0}[][]{ \shortstack{\\ $\hat{s}_0 $ }}
      \resizebox{.36\paperwidth}{!}{\includegraphics[]{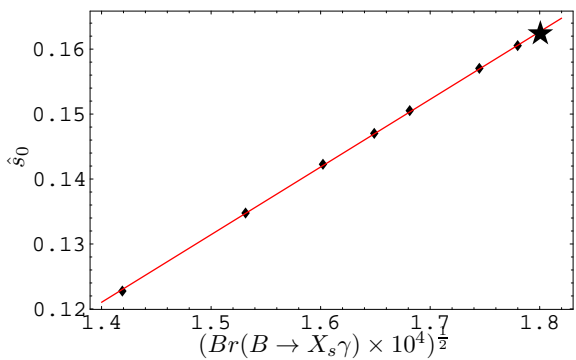}}
    
  \caption[]{\small\label{corrplot} Correlation between
    $\sqrt{Br(B\to X_s\gamma)}$  and $\hat s_0$ \cite{BPSW}. 
    The dots are the results in the ACD
    model (see Section 6)
    for $1/R = 200,250,300,350,400,600$ and $1000$ GeV  and the star
    denotes the SM value.
}
\end{figure}

{\bf 3.} It turns out that in most MFV models 
the coefficient $C_{9V}$ is only very 
weakly dependent on new physics contributions. Consequently, as pointed 
out in \cite{BPSW},
a correlation between $\hat s_0$ in $A_{\rm FB}$ and 
$Br(B\to X_s\gamma)$ exists. It is present in the ACD model discussed in 
Section 6
and in a large 
class of supersymmetric models discussed for instance in
\cite{Ali:2002jg}.  
We show this correlation in fig.~\ref{corrplot}.

\begin{figure}
\begin{center}
\includegraphics[width=9cm]{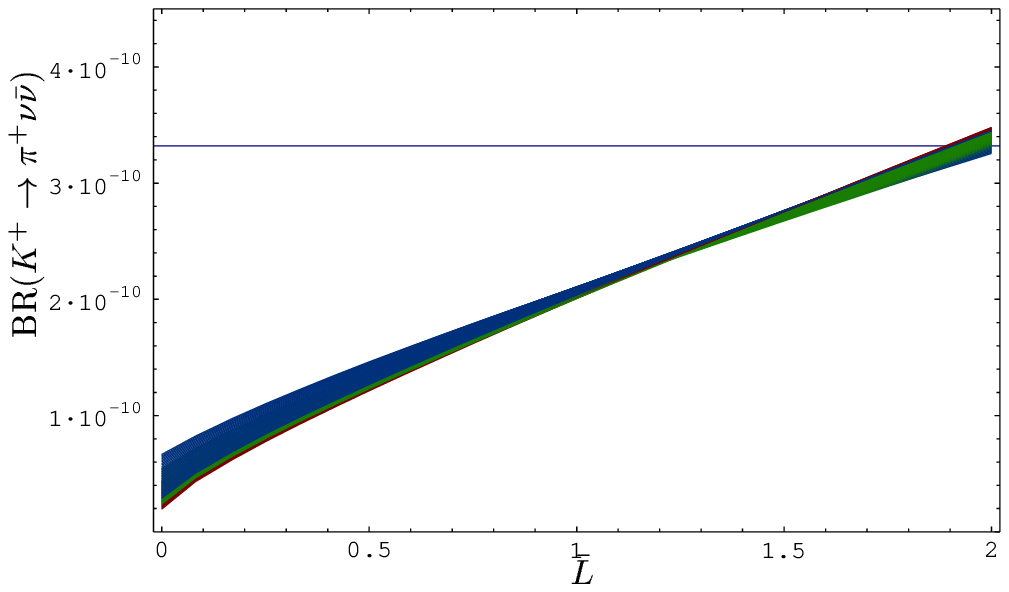}
\end{center}
\caption{Correlation between $Br(\kpn)$ and 
the $B\to\pi K$ variable $\bar L$. \label{Buras}}
\end{figure}

{\bf 4.} Very recently a correlation between the $B\to \pi K$ modes 
and $Br(\kpn)$ in a MFV new-physics scenario with enhanced $Z^0$ 
penguins has been pointed out in \cite{BFRS}. 
In fig.~\ref{Buras} we show $Br(\kpn)$ 
as a function of the variable $\bar L$ that is
given entirely in terms of $B\to \pi K$ observables and $\vub$. For a 
general discussion of correlations between  
$B\to \pi K$ decays and rare decays we refer to \cite{BFRS}.

{\bf 5.} Other correlations 
between various decays  
can be found in \cite{Buras:1998ed,Buras:1999da,Bergmann:2001pm,BF01,REL}.
For instance there exists in addition to an obvious correlation between 
$K\to\pi\nu\bar\nu$ 
and $B\to X_q\nu\bar\nu$ also a correlation between $\epe$ and rare 
semileptonic $B$ and $K$ decays.

\subsection{Model Dependent Relations}
\subsubsection{\boldmath{$B_q\to\mu\bar\mu$} and \boldmath{$\Delta M_q$}}
From (\ref{DMQMFV}) and (\ref{bbll}) we derive \cite{AJB03} 
\be\label{R2}
Br(B_{q}\to\mu\bar\mu)
=4.36\cdot 10^{-10}\frac{\tau(B_{q})}{\hat B_{q}}
\frac{Y^2(v)}{S(v)} 
\Delta M_{q}, \qquad (q=s,d).
\ee
These relations  
allow to predict $Br(B_{s,d}\to\mu\bar\mu)$  
in a given MFV model with substantially smaller hadronic uncertainties 
than found by using directly the formulae in (\ref{bbll}). In particular using
the formulae for the functions $Y$ and $S$ in the SM model of 
Section 2.5, we find \cite{AJB03}
\be\label{R3}
Br(B_{s}\to\mu\bar\mu)
=3.42\cdot 10^{-9}\left[\frac{\tau(B_{s})}{1.46~ {\rm ps}}\right]
\left[\frac{1.34}{\hat B_{s}}\right]
\left[\frac{\overline{m}_t(m_t)}{167\gev}\right]^{1.6}
\left[\frac{\Delta M_{s}}{18.0/{\rm ps}}\right], 
\ee

\be\label{R4}
Br(B_{d}\to\mu\bar\mu)
=1.00\cdot 10^{-10}\left[\frac{\tau(B_{d})}{1.54 ~{\rm ps}}\right]
\left[\frac{1.34}{\hat B_{d}}\right]
\left[\frac{\overline{m}_t(m_t)}{167\gev}\right]^{1.6}
\left[\frac{\Delta M_{d}}{0.50/{\rm ps}}\right]. 
\ee

Using $\overline{m}_t(m_t)=(167\pm 5)\gev$, the lifetimes from \cite{CERNCKM},
 $\hat B_q$ in (\ref{BBB}), $\Delta M_d=(0.503\pm0.006)/{\rm ps}$ 
 and taking as an example 
$\Delta M_s=(18.0\pm0.5)/{\rm ps}$, we find the predictions for the branching
ratios in question given in (\ref{Results}). They
 are substantially more accurate than the ones found in the 
literature in the past.

\subsubsection{\bf{$Br(\kpn)$}, \bf{$\Delta M_d/\Delta M_s$ and} 
\bf{$\beta$.} }
In \cite{BB98} an upper bound on $Br(K^+ \rightarrow \pi^+
\nu \bar{\nu})$ 
has been derived within the SM. This bound depends only on $\vcb$, $X$, 
$\xi$ and $\Delta M_d/\Delta M_s$. With the precise value for the angle 
$\beta$ now available this bound can be turned into a useful formula for 
$Br(K^+ \rightarrow \pi^+ \nu \bar{\nu})$ \cite{AI01}
that expresses 
this branching ratio in terms of theoretically clean observables. 
In any MFV model this formula reads:
\be \label{AIACD}
Br(K^+ \rightarrow \pi^+ \nu \bar{\nu})=
\bar\kappa_+\vcb^4 X^2(v)
\Bigg[ \sigma   R^2_t\sin^2\beta+
\frac{1}{\sigma}\left(R_t\cos\beta +
\frac{\lambda^4P_c(X)}{\vcb^2X(v)}\right)^2\Bigg],
\ee
where $\sigma=1/(1-\lambda^2/2)^2$, $\bar\kappa_+=7.54\cdot 10^{-6}$, 
$P_c(X)=0.39\pm 0.06$ and $R_t$ is given in (\ref{Rt}).
This formula is theoretically very clean and does not involve 
hadronic uncertainties except for $\xi$  and to a lesser 
extent in $\vcb$. We will use it in Section 6.
\subsubsection{\boldmath{$\kpn$}, \boldmath{$\klpn$} and 
\boldmath{$\kmm$}}
From (\ref{bkpn}), (\ref{bklpn}) and (\ref{bklm}) it is possible to derive 
a relation between the three decays in question \cite{BH92} 
$(\bar P_c=P_c(1-\lambda^2/2))$
\be\label{R8}
B_1=B_2+\left[(\bar P_c(Y)+\sqrt{B_3})\frac{X(v)}{Y(v)}-\bar P_c(X)\right]^2,
\qquad 
B_3={Br(\kmm)_{\rm SD}\over 1.95\cdot 10^{-9}}
\ee
with $B_1$ and $B_2$ defined in (\ref{b1b2}).
Consequently
\be
\frac{X(v)}{Y(v)}=\frac{\bar P_c(X)+\sqrt{B_1-B_2}}{\bar P_c(Y)+\sqrt{B_3}}.
\ee
where the signs in case of ambiguities have been chosen as in the SM.

\section{Procedures for the Determination of Master Functions}
\subsection{Preliminaries}
The idea to determine the values of the master functions in a model 
independent manner is not new. The first model independent determination of 
$S(v)$ has been presented in \cite{BCKO}, subsequently in \cite{Santamaria}
 and very recently 
in \cite{BUPAST}. The 
corresponding analyses for $X(v)$ and $Y(v)$ can be found in 
\cite{BF01,Bergmann:2001pm} and 
\cite{AMGIISST,BHI}, 
respectively. To my knowledge, no direct determination of the remaining four 
functions can be found in the literature, but the functions $Z(v)$, $D'(v)$
 and $E'(v)$ 
can be in principle extracted from the model independent analyses that use 
the Wilson coefficients instead of master functions \cite{Ali:2002jg}. 
The function $E(v)$
 is very 
difficult to determine as we will see below.

In what follows we will assume that the UUT has been found so that the 
universal CKM matrix is known. Moreover, we will assume first that all
non-perturbative factors like $\hat B_K$, $F_{B_q}$ have been calculated with 
sufficient precision. Subsequently we will relax this assumption.
\subsection{Ideal Scenario}
With the assumptions just made, it is straightforward to determine the 
seven master functions and simultaneously test the general idea of MFV. 
Here we go:

{\bf Step 1:}

$S(v)$ can be extracted from $\varepsilon_K$, $\Delta M_d$ and $\Delta M_s$. 
Finding the same value of $S(v)$ in these three cases, would be a significant 
sign for the dominance of MFV in $\Delta F=2$ transitions.

{\bf Step 2:}

$X(v)$ can be extracted from $\kpnn$, $\klpn$, $B\to X_s\nu\bar\nu$ and 
$B\to X_d\nu\bar\nu$. Again, if MFV is the whole story in the $\Delta F=1$ 
decays with $\nu\bar\nu$ in the final state, the same  
value of $X(v)$ should result in these four cases. 

{\bf Step 3:}

$Y(v)$ can be extracted from the short distance dispersive part of 
$K_L\to\mu^+\mu^-$, the parity violating asymmetry $\Delta_{\rm LR}$ 
and the branching ratios 
$B_{d,s}\to\mu^+\mu^-$. The comments are as in Step 2 with 
$\nu\bar\nu$ replaced by $\mu\bar\mu$.

{\bf Step 4:}

Having determined $X(v)$ and $Y(v)$, we can extract the values of 
$Z(v)$ and $E(v)$ by studying simultaneously $\kpe$ and $\epe$.

{\bf Step 5:}

With all this information at hand we can finally find the values of 
$D'(v)$ and $E'(v)$ from a combined analysis of $B\to X_s\gamma$, 
   $B\to X_s~{\rm gluon}$  and 
$B\to X_s\mu^+\mu^-$. To this end the branching ratio for
$B\to X_s\gamma$ and the forward-backward asymmetry $A_{\rm FB}$ in
$B\to X_s\mu^+\mu^-$ are most suitable.

This scenario is rather unrealistic for the present decade as the 
theoretical uncertainties in several of the observables are presently 
large and their precise measurements will still take some time. 
This is in particular the case for $\kmm$, $\kpe$, $B\to X_s~{\rm gluon}$ 
and $\epe$. 
Let us 
then investigate, whether by making plausible assumptions about the 
importance of various master functions, we can determine the values of 
all the relevant functions using only observables that are theoretically 
rather clean.

\subsection{Realistic Scenario}
The idea here is to assume as in Section 2.8 that the dominant new physics 
contributions 
reside only in five functions $C(v)$, $Z(v)$, $S(v)$, $D'(v)$ and $E'(v)$ 
and to set 
the remaining ones to the SM values.
Here we go again but in a somewhat different order:

{\bf Step 1:}

We determine $X(v)$ from $\kpnn$, $\klpn$, $B\to X_s\nu\bar\nu$ and 
$B\to X_d\nu\bar\nu$ as in the Step 2 of the previous scenario. 
Taking $B^{\nu\bar\nu}$ from the SM, we can determine $C(v)$.
All
these decays are theoretically clean and the success of this determination 
is in the hands of the experimentalists and their sponsors.

{\bf Step 2:}

The knowledge of $C(v)$ together with $B^{\mu\bar\mu}$ from the SM,
allows to determine $Y(v)$. Similar we could take $D(v)$ from the SM 
to find the value of $Z(v)$ but this is not necessary as seen in the 
Step 5 below.

{\bf Step 3:}

From the ratio $Br(B_s\to \mu^+\mu^-)/\Delta M_s$, that is theoretically 
rather clean, we can determine the value of $Y^2(v)/S(v)$ 
by means of (\ref{R2})
and consequently 
$S(v)$. 

{\bf Step 4:}

Neglecting the small contribution of
$E'(v)$ to $B\to X_s\gamma$ we can determine 
$D'(v)$.

{\bf Step 5:}

Having $Y(v)$ and $D'(v)$ at hand we can next extract $Z(v)$ from 
$A_{\rm FB}$, in particular from the value of $\hat s_0$.

$E(v)$ and $E'(v)$ could then be determined from $\epe$ and 
$B\to X_s~{\rm gluon}$, respectively.
 However, these determinations 
are rather unrealistic in view of the subdominant role of $E(v)$ in $\epe$, 
 large hadronic uncertainties in $P_r$ in (\ref{FE}), very large
renormalization scheme dependence in $Br(B\to X_s~{\rm gluon})$ 
and great difficulty in extracting this branching ratio from the data.

\subsection{Sign Ambiguities}
Needless to say, in the procedures outlined above we did not discuss
the sign ambiguities in the determination of master functions from branching 
ratios. These 
ambiguities can easily be resolved when several quantities are considered 
simultaneously. For instance while $\klpn$ and $B\to X_s\nu\bar\nu$ are not
sentsitive to the sign of $X(v)$, $Br(\kpn)$ is substantially smaller for 
negative $X(v)$. Similar $A_{\rm FB}$, $Br(B\to X_s l^+l^-)$, $\kmm$, 
$\Delta_{\rm LR}$, $\kpe$ and $\epe$ are sensitive to the signs of 
the master functions. 
These aspects have been already partially investigated in \cite{BF01,BFRS} and 
it will be of interest to return to them in the future when more data are 
available.

\subsection{An Example}
In view of  limited data on FCNC processes a numerical analysis along 
the steps suggested above will not be done here. Instead we will present an 
example. To this end let note that 
from the analyses in \cite{BUPAST,Ali:2002jg,BFRS} one can infer the 
upper bounds $S(v)\le 3.8$ and 
\be\label{XYZb}
X(v)\le 2.7, \qquad Y(v)\le 2.2, \qquad Z(v)\le 1.9
\ee
to which I would not like to attach any confidence level. The  
bounds in (\ref{XYZb}) follow from the Belle and BaBar data 
\cite{Kaneko:2002mr} on $B\to X_s l^+l^-$ under the assumption that 
the new physics contributions to $\Delta F=1$ and the function $D$ can be 
neglected \cite{BFRS}. Let us then calculate the values of various 
branching ratios assuming the maximal values for the functions $X$, $Y$ and 
$Z$ in (\ref{XYZb}). 

The CKM parameters are fixed in the following manner. We set $\lambda=0.224$,
$\vcb=0.0415$ and $\sin\beta=0.40$. We next assume that $\Delta M_s=18.0/ps$ 
and use (\ref{Rt}) with $\xi=1.24$ to find UUT with (see (\ref{UUTR}))
\be\label{UUTOUT}
R_t=0.91, \quad R_b=0.40, \quad \bar\varrho=0.166, \quad
\bar\eta=0.364,\quad \frac{\vts}{\vcb}=0.983,
\ee
\be\label{UUTOUT2}
\vtd=8.55\cdot 10^{-3},
\qquad \IM\lambda_t=1.44\cdot 10^{-4}, \qquad 
\RE\lambda_t=-3.14\cdot 10^{-4}.  
\ee
\begin{table*}[hbt]
\vspace{0.4cm}
\begin{center}
\caption[]{\small \label{brMFV} Example of branching ratios for rare decays 
in the MFV and  the SM.
}
\begin{tabular}{|c||c|c|}
\hline
{Branching Ratios} &  MFV &   SM
 \\ \hline
$Br(\kpn)\times 10^{11}$ &  $19.1$ &  $8.0$ 
\\ \hline
$Br(\klpn)\times 10^{11}$ &  $9.9$ &   $3.2$ 
\\ \hline
$Br(\kmm)_{\rm SD}\times 10^{9} $ &  $3.5$ &  $0.9$
\\ \hline
$Br(K_{\rm L} \to \pi^0 e^+ e^-)_{\rm CPV}\times 10^{11}$ &  $4.9$ & $3.2$ 
\\ \hline
$Br(B\to X_s\nu\bar\nu)\times 10^{5}$ &  $11.1$ & $3.6$ 
\\ \hline
$Br(B\to X_d\nu\bar\nu)\times 10^{6}$ &  $4.9$ & $1.6$ 
\\ \hline
$Br(B_s\to \mu^+\mu^-)\times 10^{9}$ &  $19.4$ &  $3.9$ 
\\ \hline
$Br(B_d\to \mu^+\mu^-)\times 10^{10}$ &  $6.1$ & $1.2$ 
\\ \hline
\end{tabular}
\end{center}
\end{table*}

The result of this exercise is shown in column MFV of
table~\ref{brMFV} where also 
the SM results with $X=1.53$, $Y=0.98$ and $Z=0.68$ are shown.
In the case of $B_q\to \mu^+\mu^-$ we have used central values 
of $F_{B_q}$ and $\tau(B_q)$ \cite{CERNCKM}.
While somewhat higher values of branching ratios can still be obtained when 
the input parameters are varied, this exercise shows that enhancements 
of branching ratios in MFV by more than factors of six relative to the SM 
should not be expected. Of interest is the high value of
$Br(\kmm)_{\rm SD}$. It indicates that this decay could give a strong 
upper bound on the function $Y$ if the hadronic uncertainties could be put 
under control \cite{AMGIISST}. 
Even a better example is $\epe$ \cite{Buras:1998ed}. 
 With the matrix elements in (\ref{PI1}) we could get  
very strong upper bounds on the functions 
$X$, $Y$ and $Z$ as with the values in (\ref{XYZb}) we find 
$\epe=-2.4 \cdot 10^{-4}$ in total disagreement with the experimental 
data in (\ref{eps}). In the SM we find $\epe=17.4 \cdot 10^{-4}$. 
Yet, with  the matrix elements in (\ref{PI2}) the MFV scenario considered 
here gives $\epe=16.1\cdot 10^{-4}$ in perfect agreement with (\ref{eps}).
This example demonstrates that until the values of $P_r$ in $\epe$ are 
put under control, this ratio cannot be efficiently used as a constraint 
on MFV models. 

A similar analysis in a different spirit and a different set of input 
parameters prior to the data of \cite{Kaneko:2002mr} can be found in 
\cite{AMGIISST}.

\section{MFV and Universal Extra Dimensions}

\subsection{Introduction}
Let us next discuss the master functions and their phenomenological 
implications in a specific MFV model: the SM model with one extra 
universal dimension. This is the model due to Appelquist, Cheng and 
Dobrescu (ACD) \cite{Appelquist:2000nn} in which all the SM fields 
are allowed to propagate in all available dimensions.  
In this model the relevant 
penguin and box diagrams receive additional contributions from 
Kaluza-Klein (KK) modes and from the point of view of FCNC 
processes the only additional free parameter relative to the SM is the
compactification scale $1/R$.  
Extensive analyses of the precision electroweak data, the analyses of the 
anomalous magnetic moment
of the muon and of the $Z\to b\bar b$ vertex have shown the consistency of the 
ACD model with the data for $1/R\ge 250\gev$.
We refer to \cite{BSW02,BPSW} for the relevant papers.

The question then arises whether such low compactification scales are still 
consistent with the data on FCNC processes. This question has been addressed 
in detail in \cite{BSW02,BPSW,durham}. The answer is given below.

\subsection{Master Functions in the ACD Model}
The master functions in the ACD model  become functions 
of $x_t$ and $1/R$: $F_r(x_t,1/R)$. 
They have been calculated in 
\cite{BSW02,BPSW}  with the results given
in table~\ref{inamitab}. 
Our results for the function $S$ have been confirmed in
\cite{Chakraverty:2002qk}. 
For $1/R=300~\gev$, the functions $S$, $X$, $Y$, $Z$ 
are enhanced by $8\%$, $10\%$, $15\%$ and $23\%$ relative to the SM values, 
respectively. The impact of the KK modes on the function $D$  and 
the $\Delta F=1$ box functions is negligible in accordance with 
our assumptions in subsection 2.8.

The most interesting 
are very strong suppressions of $D'$ and $E'$, that for $1/R=300\gev$ amount 
to $36\%$ and $66\%$ relative to the SM values, respectively. 
However, the  effect of these suppressions  is softened in 
the relevant branching ratios through sizable additive QCD corrections, 
discussed already in Section 4.

\begin{table*}[hbt]
\begin{center}
\caption[]{\small \label{inamitab} Values for the functions $S$, $X$, $Y$, 
$Z$, $E$, $D'$, $E'$, $C$ and $D$.
}
\begin{tabular}{|c||c|c|c|c|c|c|c||c|c|}\hline
 $1/R~[{\rm GeV}]$  & {$S$} & {$X$}& {$Y$} & {$Z$} & {$E$} & {$D'$} & {$E'$} 
& {$C$} & $D$
 \\ \hline
$ 250$ & $ 2.66 $ &  $1.73 $ &  $1.19$ & $0.89$ & $0.33$  & $0.19$ &$ 0.02$ &
$1.00$  & $-0.47$
\\ \hline
$300$  & $ 2.58 $ &  $1.67 $ &  $1.13$ & $0.84$ & $0.32$  & $0.24$  &$ 0.07$
&  $0.95$  & $-0.47$
\\ \hline
$400$  & $ 2.50 $ &  $1.61 $ &  $1.07$ & $0.77$ & $0.30$  & $0.30$  &$ 0.12 $
& $0.89$  & $-0.47$
\\ \hline
SM     & $2.40$   &  $ 1.53 $ & $0.98$ & $0.68$ & $0.27$  &$ 0.38$ & $ 0.19$ 
& $0.80$  & $-0.48$
\\ \hline
\end{tabular}
\end{center}
\end{table*}

\subsection{The Impact of the KK Modes on Specific Decays}
\subsubsection{The Impact on the Unitarity Triangle}
Here the function $S$ plays the crucial role. Consequently the impact 
of the KK modes on the UT is rather small. For $1/R=300\gev$, $\vtd$, 
$\bar\eta$ and $\gamma$ are suppressed by $4\%$, $5\%$ and $5^\circ$, 
respectively. It will be difficult to see these effects in the 
$(\bar\varrho,\bar\eta)$ plane. On the other hand a $4\%$ suppression 
of $\vtd$ means a $8\%$ suppression of the relevant branching ratio for a
rare decay sensitive to $\vtd$ and this effect has to be taken into account. 
Similar comments apply to $\bar\eta$ and $\gamma$. 
As we work now in a specific model, we follow here a different philosophy
than in the model independent analysis of the previous sections and 
determine the CKM parameters using also the $S$ function, like in the 
analysis of Section 3.

\subsubsection{The Impact on Rare K and B decays}
Here the dominant KK effects enter through the function $C$ or equivalently
$X$ and $Y$, depending on the decay considered. In table~\ref{brtable} 
we show seven branching ratios as functions of $1/R$ for central values of 
all remaining input parameters. 
For $1/R=300\gev$ the following enhancements relative to the SM predictions 
are seen:
 $\kpn~(9\%)$, $\klpn~(10\%)$, 
$B\to X_{d}\nu\bar\nu~(12\%)$, $B\to X_{s}\nu\bar\nu~(21\%)$, 
$K_L\to\mu\bar\mu~(20\%)$, $B_{d}\to\mu\bar\mu~(23\%)$ and 
$B_{s}\to\mu\bar\mu~(33\%)$.
The SM values in table~\ref{brtable} differ slightly from those given in 
the example of table~\ref{brMFV} due to a different choice of the CKM 
parameters in \cite{BSW02}.
\begin{table*}[hbt]
\vspace{0.4cm}
\begin{center}
\caption[]{\small Branching ratios for rare decays in the ACD model and the 
SM.
\label{brtable}}
\begin{tabular}{|c||c|c|c|c|}
\hline
{$1/R$ } &  {$250\gev$}&  {$300\gev$} 
& {$400\gev$} & SM
 \\ \hline
$Br(\kpn)\times 10^{11}$ &  $8.36$ & $ 8.13$ & $ 7.88$ &  $7.49$ 
\\ \hline
$Br(\klpn)\times 10^{11}$ &  $3.17$ & $ 3.09 $ & $ 2.98$ &  $2.80$ 
\\ \hline
$Br(\kmm)_{\rm SD}\times 10^{9} $ &  $1.00$ &  $ 0.95$ & $ 0.88$ &
$0.79$\\ \hline
$Br(B\to X_s\nu\bar\nu)\times 10^{5}$ &  $4.56$ & $ 4.26 $ & $ 3.95$ &
$3.53$ \\ \hline
$Br(B\to X_d\nu\bar\nu)\times 10^{6}$ &  $1.70$ & $ 1.64$ & $ 1.58$ &
$1.47$ \\ \hline
$Br(B_s\to \mu^+\mu^-)\times 10^{9}$ &  $5.28$ & $ 4.78$ & $4.27$ &
$3.59$ \\ \hline
$Br(B_d\to \mu^+\mu^-)\times 10^{10}$ &  $1.41$ & $ 1.32 $ & $ 1.22$ &
$1.07$ \\ \hline
\end{tabular}
\end{center}
\end{table*}

\subsubsection{An Upper Bound on \bf{$Br(\kpn)$} in the ACD Model}
The enhancement of $Br(\kpn)$ in the ACD model is interesting in view
of the results from the BNL E787
collaboration \cite{Adler97} in (\ref{kp01}) 
with the central value  by a factor of 2 above the SM expectation. Even if 
the errors are substantial and this result is compatible with the SM, 
the ACD model with a low compactification scale is  closer to the 
data.

In order to find the upper bound on $Br(\kpn)$ in the ACD model we use
the formula (\ref{AIACD}) with $X(v)$ given in table~\ref{inamitab} and  
$\vcb\le 0.0422$, $P_c(X)<0.47$, $\sin\beta=0.40$ and $\mt<172~\gev$.
Here we have set $\sin 2 \beta = 0.734$, its central value, as 
$Br(\kpn)$ depends very weakly on it.  
The result of this exercise is shown in table~\ref{Bound}. We give 
there $Br(\kpn)_{\rm max}$  
as a function of $\xi$ and $1/R$ for two different values of $\Delta M_s$.
We observe that for $1/R=250~\gev$ and $\xi=1.30$ the maximal value
for $Br(\kpn)$ in the ACD model is rather close to the central value in
(\ref{kp01}). 

\begin{table*}[hbt]
\vspace{0.4cm}
\begin{center}
\caption[]{\small Upper bound on $Br(\kpnn)$ in units of $10^{-11}$ for 
different 
values of $\xi$, $1/R$ and $\Delta M_s=18/{\rm ps}~(21/{\rm ps})$ from 
\cite{BSW02}.  
\label{Bound}}
\begin{tabular}{|c||c|c|c|c|}\hline
{$\xi$ } & {$1/R=250~\gev$}& {$1/R=300~\gev$} 
& {$1/R=400~\gev$} & SM
 \\ \hline
$1.30$ &  $ 12.7^*~(11.3^*)$ & $ 12.0^*~(10.7)$ & 
$ 11.3^*~(10.1)$ &  $10.8~(9.3)$ \\ \hline
$1.25$ &  $ 12.0~(10.7) $ & $ 11.4~(10.2) $ & 
$ 10.7~(9.6)$ & $10.3~(8.8)$ \\ \hline
$1.20$ & $ 11.3~(10.1) $ & $ 10.7~(9.6) $ & 
$ 10.1~(9.1)$ & $9.7~(8.4)$\\ \hline
$1.15$ &  $10.6~(9.5) $ & $ 10.1~(9.0) $ & $ 9.5~(8.5)$ &
$9.1~(7.9)$ \\ \hline
\end{tabular}
\end{center}
\end{table*}

\subsubsection{The Impact on \bf{$B\to X_s\gamma$} and
   \bf{$B\to X_s~{\rm gluon}$}}

Due to strong suppressions of the functions $D'$ and $E'$ by the KK modes, 
the $B\to X_s\gamma$ and $B\to X_s~{\rm gluon}$ decays are considerably 
suppressed compared to SM estimates. For $1/R=300\gev$, 
$Br(B\to X_s\gamma)$ is suppressed by $20\%$, while $Br(B\to X_s~{\rm gluon})$
even by $40\%$. The phenomenological relevance of the latter suppression is 
unclear at present as $Br(B\to X_s~{\rm gluon})$ suffers from large 
theoretical uncertainties and its extraction from experiment is very
difficult. The suppression of $Br(B\to X_s\gamma)$ in the ACD model has 
already been found in an approximate calculation of \cite{Agashe:2001xt}.

\begin{figure}[]
\renewcommand{\thesubfigure}{\space(\alph{subfigure})}
\centering
\psfragscanon

  \psfrag{bsgammabsgammabsg}{ $Br(B\rightarrow X_s \gamma )\times 10^{4}$}
  \psfrag{rinvrinv}[][]{ \shortstack{\\  $R^{-1}$ [GeV] }}
  \label{bsg.eps}
        \resizebox{.38\paperwidth}{!}{\includegraphics[]{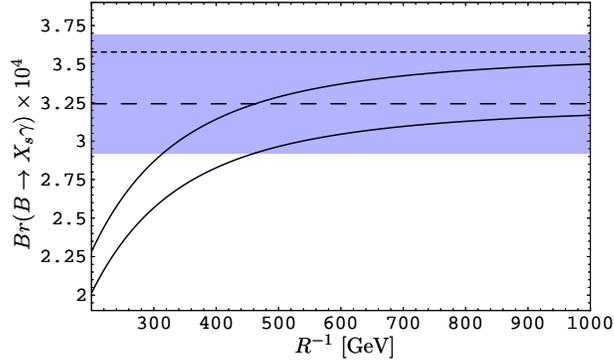}}
    \caption[]{\small\label{bsgplot} The branching ratio for $B\to
      X_s\gamma$  as a function of $1/R$. 
See text.}
  \end{figure}

In fig.~\ref{bsgplot} we compare $Br(B\to X_s\gamma)$ 
in the ACD model with the experimental data and with the expectations 
of the SM. The shaded region represents the data in (\ref{bsgexp}) and the 
upper (lower) dashed horizontal line are the central values in the SM 
for $m_c/m_b=0.22~(m_c/m_b=0.29)$. The solid lines represent the 
corresponding central values in the ACD model. The theoretical errors, 
not shown in the plot, are for all curves roughly $\pm 10\%$

We observe that in view 
of the sizable experimental error and considerable parametric uncertainties in 
the theoretical prediction, the strong suppression of $Br(B\to X_s\gamma)$ 
by the KK modes does not yet provide a powerful lower bound on $1/R$ and the
values $1/R\ge 250\gev$ are fully consistent with the experimental result. 
Once the uncertainty due to $m_c/m_b$  and the experimental uncertainties 
are reduced, 
$Br(B\to X_s\gamma)$ may provide a very powerful bound on $1/R$ that is 
substantially stronger than the bounds obtained from the electroweak precision 
data.

\subsubsection{The Impact on  \bf{$A_{FB}(\hat s)$}}
\begin{figure}[hbt]
\renewcommand{\thesubfigure}{\space(\alph{subfigure})} 
  \centering 
  \subfigure[]{\psfragscanon
  \psfrag{nfbnfb}{ $\hat{A}_{FB}$}
  \psfrag{hats}[][]{ \shortstack{\\ $\hat{s} $ }}
      \resizebox{.28\paperwidth}{!}{\includegraphics[]{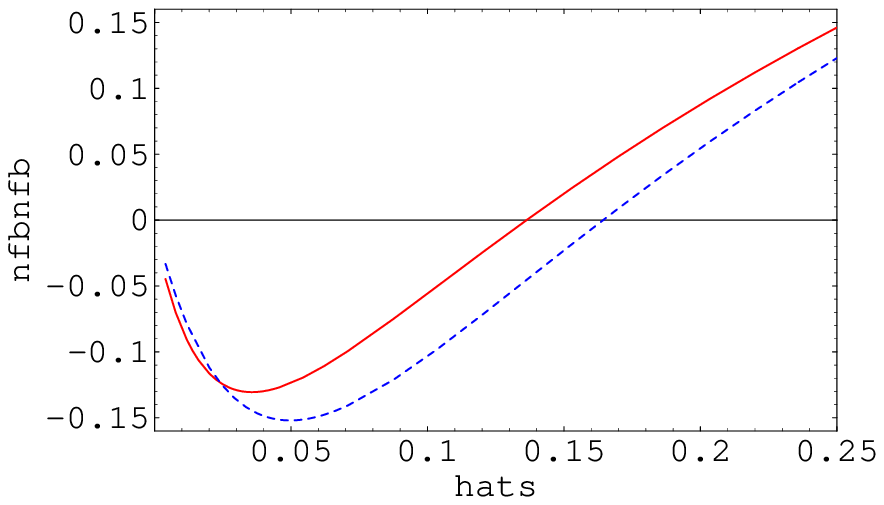}}
    }
  \hspace{0.1cm}
  \subfigure[]{\psfragscanon
  \psfrag{zafb}{ $\hat{s}_{0}$}
  \psfrag{rinvrinv}[][]{ \shortstack{\\  $R^{-1}$ [GeV]  }}
    \resizebox{.28\paperwidth}{!}{ \includegraphics[]{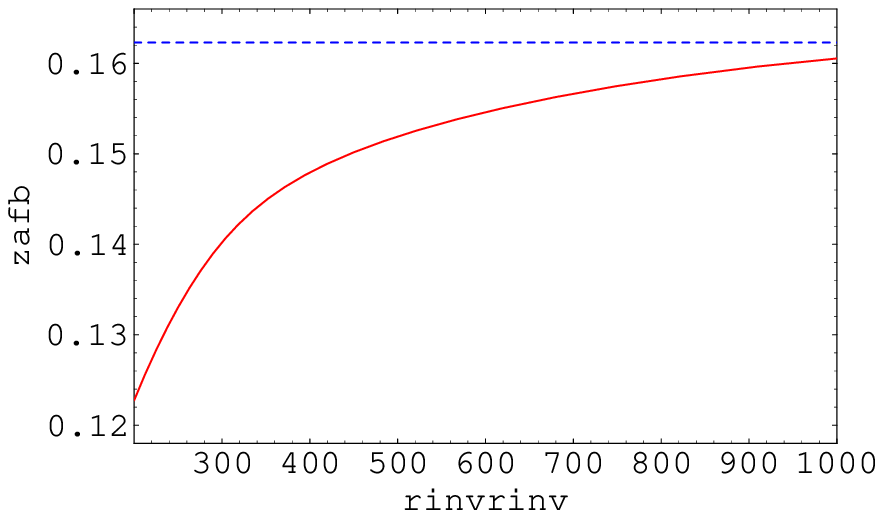}}
    }
  \caption[]{\small\label{normalizedfb} (a) Normalized Forward-Backward
    asymmetry in the SM (dashed line) and ACD for $R^{-1}=250$ GeV. 
(b) Zero of the forward backward
    asymmetry $A_{FB}$.}
\end{figure}

In fig.~\ref{normalizedfb} (a)  we show the normalized
Forward-Backward asymmetry, given in (\ref{ABF}), for $1/R=250\gev$. 
The dependence of 
$\hat s_0$ on $1/R$  
is shown in fig.~\ref{normalizedfb} (b). 
We observe that the value of $\hat s_0$ is considerably reduced relative 
to the SM result obtained by including NNLO corrections 
\cite{Ali:2002jg,NNLO1,NNLO2}. This decrease, as seen in fig.~\ref{corrplot}, 
is related to the decrease 
of $Br(B\to X_s\gamma)$.
For 
$1/R=300\gev$ we find the value for $\hat s_0$ that is  very close to 
the NLO prediction of the 
SM. This result demonstrates very clearly the importance of the
calculations of the higher 
order QCD corrections, in particular in quantities like $\hat s_0$
that are theoretically clean. We expect that the results in
figs.~\ref{normalizedfb} (a) and (b) 
will play an important role in the tests of the ACD model in the future.

\subsection{Concluding Remarks}
The analysis of the ACD model discussed above shows that all the present 
data on FCNC 
processes are consistent with $1/R$ as low as $250\gev$, implying that 
the KK particles could in principle be found already at the Tevatron.
Possibly, the most interesting results of our analysis is the 
enhancement of $Br(\kpn)$, the sizable downward 
shift of  the zero ($\hat s_0$) in the 
$A_{\rm FB}$ asymmetry and the suppression of $Br(B\to X_s\gamma)$.

The nice feature of this extension of the SM is the presence of only one 
additional parameter, the compactification scale. This feature allows a
unique pattern of various
enhancements and suppressions relative to the SM expectations. 
We would like to emphasize that violation of this pattern by the future 
data will
exclude the ACD model. For instance a measurement of $\hat s_0$ that is 
higher than the SM estimate would automatically exclude this model as 
there is no compactification scale for which this could be satisfied.
Whether these enhancements and suppressions are required by the data or 
whether they exclude the ACD model with a low compactification scale, 
will depend 
on the precision of the forthcoming experiments and the efforts to decrease 
the theoretical uncertainties.

\section{Summary}
In these lectures we have discussed the class of models with MFV as defined 
in \cite{UUT}. See Section 2 for details. While 
these models, including the SM, were with us already for many years,  
we are only beginning to put them under decisive tests. We have emphasized 
here the parametrization of MFV models in terms of master functions 
\cite{PBE,BH92} that has been already efficiently used within the SM since 
1990. This should be contrasted with the formulation
in terms of 
Wilson coefficients of certain local operators as done for instance in
\cite{ALIMFV,Ali:2002jg}. 
While the latter formulation involves scales as low as $\ord(m_b)$
 and $\ord(1\gev)$, the 
former one exhibits more transparently the short distance contributions at 
scales $\ord(M_W,m_t)$ and higher. 
The formulation presented here has also the advantage that 
it allows to formulate the $B$ and $K$ decays in terms of the same building 
blocks, the master functions. This allows to study transparently the 
correlations between not only different $B$ or $K$ decays but also between 
$B$ and 
$K$ decays. This clearly is much harder when working directly with the Wilson 
coefficients. In particular the study of the correlations between $K$ and $B$
decays is very difficult as the Wilson coefficients in $K$ and $B$ decays, 
with 
a few exceptions, involve different renormalization scales. 
 I believe that the 
present formulation will be more useful when the data on all relevant decays 
will be available.

We have also compared briefly our definition of MFV to a slightly
more general MFV framework developed in \cite{AMGIISST}. This comparison 
is given in Section 2.3.
In particular in the latter framework the new physics contributions cannot 
be always taken into account by simply modifying the master functions as 
done in our approach. As a result some of the correlations present in our
approach are not present there. Only time will show which of these two 
frameworks gives a better description of the data. 

We have outlined various procedures for a model independent determination 
of the master functions. While such an approach cannot replace the direct 
calculation of master functions in a given model, it may help to test the 
general concept of MFV. In this context of particular interest are relations 
between various observables that do not involve the master functions at all. 
These relations, if violated, would imply new sources of flavour violation 
without the need for precise knowledge of $F_r$.

At present most of the experimental data that we have to our disposal, are 
consistent with MFV but this information is still rather limited. On the other 
hand there are at least two pieces of data that could point towards 
the importance of  new operators, new 
sources of flavour violation and in particular of CP violation.

These are:
\begin{itemize}
\item
The violation of the second relation in (\ref{R7}) as seen by Belle 
\cite{Bela} but not BaBar \cite{Bab}. A subset of papers discussing this 
issue is given in \cite{PHIKS}.     
\item
 The $B\to \pi K$  puzzle: $\bar L$ in fig.~\ref{Buras} 
is required by the $B\to \pi K$ data to be larger than $1.8$, 
whereas the MFV correlation between 
$B\to\pi K$ and $B\to X_s\mu^+\mu^-$ indicates $\bar L <1.1$ \cite{BFRS}. 
If confirmed with higher precision, this result would 
imply physics beyond  MFV.
\end{itemize}
Only time will show whether these findings will require to go beyond MFV.

We have also presented the results of an explicit calculation of master 
functions in the SM with one additional universal extra dimension. The 
dependence of $F_r$ on a single new parameter, the compactification radius 
$R$, allowed to predict uniquely the signs of the KK contribution to 
$F_r$. This will 
allow in the future to test this model when $F_r$ will be extracted from the 
data.

Our story of the Minimal Flavour Violation is approaching the end. I hope I 
have convinced some readers that this framework is a very good starting 
point for going beyond the Standard Model. We should know already in the 
coming years
whether indeed the description of all FCNC processes in terms of seven 
or even only four master functions will survive 
all future tests. In this spirit it will be exciting to follow the
experimental developments and to see whether all the MFV correlations 
between various observables are confirmed by the future experimental 
findings.

{\bf Acknowledgements}

I would like to thank the organizers for inviting me to such a 
wonderful school  and most enjoyable atmosphere. 
I also thank Stefan Recksiegel, Felix Schwab and Andreas Weiler for 
comments on the manuscript and Frank Kr\"uger for useful discussions.
The work presented here has been supported in part by the German 
Bundesministerium f\"ur
Bildung und Forschung under the contract 05HT1WOA3 and the 
DFG Project Bu. 706/1-2.

\vfill\eject


\begin{thebibliography}{80}
\bibitem{WARN} 
The numerical value on the r.h.s of (\ref{100}) being proportional to
$\lambda^{-10}$ is very sensitive to $\lambda$ but as $A\propto \lambda^{-2}$
and $P_c(\varepsilon)\propto \lambda^{-4}$ the constraint (\ref{100}) depends
only quadratically on $\lambda$.
\end{thebibliography}
\end{document}